\documentclass{aa}
\usepackage[varg]{txfonts}
\usepackage[colorlinks=true,linkcolor=blue,citecolor=blue]{hyperref}
\usepackage{graphicx}
\usepackage{amsmath,amsfonts,amssymb,mathrsfs}
\usepackage{aas_macros}
\usepackage{scalerel,stackengine}
\stackMath

\usepackage[dvipsnames]{xcolor}

\newcommand\reallywidehat[1]{%
\savestack{\tmpbox}{\stretchto{%
  \scaleto{%
    \scalerel*[\widthof{\ensuremath{#1}}]{\kern-.6pt\bigwedge\kern-.6pt}%
    {\rule[-\textheight/2]{1ex}{\textheight}}
  }{\textheight}%
}{0.5ex}}%
\stackon[1pt]{#1}{\tmpbox}%
}
\begin{document} 

 \title{Redundant apodization for direct imaging of exoplanets 2: \\ Application to island effects}
 \titlerunning{Redundant apodization for direct imaging of exoplanets 2}
\authorrunning{Leboulleux et al.}
  \author{Lucie Leboulleux \inst{1}, Alexis Carlotti \inst{1}, Mamadou N'Diaye \inst{2}, Arielle Bertrou-Cantou \inst{3,4}, Julien Milli \inst{1}, Nicolas Pourré \inst{1}, Faustine Cantalloube \inst{5}, David Mouillet \inst{1}, Christophe Vérinaud \inst{6}}
  \institute{Univ. Grenoble Alpes, CNRS, IPAG, 38000 Grenoble, France
  \and Université Côte d’Azur, Observatoire de la Côte d’Azur, CNRS, Laboratoire Lagrange, France
  \and LESIA, Observatoire de Paris, Université PSL, CNRS, Sorbonne Université, Université de Paris, 5 place Jules Janssen, 92195 Meudon, France
  \and Department of Astronomy, California Institute of Technology, Pasadena, CA 91125, USA
  \and Aix Marseille Université, CNRS, LAM (Laboratoire d’Astrophysique de Marseille) UMR 7326, 13388 Marseille, France
  \and European Southern Observatory (ESO), Karl-Schwarzschild-Str. 2, D-85748 Garching bei Muenchen, Germany
  \\
             \email{lucie.leboulleux@univ-grenoble-alpes.fr}} 

  \abstract
   {Telescope pupil fragmentation from spiders generates specific aberrations that have been observed at various telescopes and are expected on the 30+-meter class telescopes under construction. This so-called island effect induces differential pistons, tips and tilts on the pupil petals, deforming the instrumental point spread function (PSF), and is one of the main limitation to the direct detection of exoplanets with high-contrast imaging. These petal-level aberrations can have different origins such as the low-wind effect or petaling errors in the adaptive optics reconstruction.}
   {In this paper, we propose to alleviate the impact of the aberrations induced by island effects on high-contrast imaging by adapting the coronagraph design in order to increase its robustness to petal-level aberrations.}
   {Following a method first developed and applied on robustness to errors due to primary mirror segmentation (segment phasing errors, missing segments...), we develop and test Redundant Apodized Pupils (RAP), i.e. apodizers designed at the petal-scale, then duplicated and rotated to mimic the pupil petal geometry.}
   {We apply this concept to the ELT architecture, made of six identical petals, to yield a $10^{-6}$ contrast in a dark region from $8$ to $40\lambda/D$. Both amplitude and phase apodizers proposed in this paper are robust to differential pistons between petals, with minimal degradation to their coronagraphic PSFs and contrast levels. In addition, they are also more robust to petal-level tip-tilt errors than classical apodizers designed for the whole pupil, with which the limit of contrast of $10^{-6}$ in the coronagraph dark zone is achieved for constraints up to $2$ rad RMS of these petal-level modes.}
   {In this paper, the RAP concept proves its robustness to island effects (low-wind effect and post-adaptive optics petaling), with an application to the ELT architecture. It can also be considered for other 8- to 30-meter class ground-based units such as VLT/SPHERE, Subaru/SCExAO, GMT/GMagAO-X, or TMT/PSI.}
    

   \maketitle

\section{Introduction}
\label{s:Introduction}

The direct imaging and spectroscopy of exoplanets has pushed forward the development of high-contrast imagers set up on large telescopes: the Spectro-Polarimetric High-contrast Exoplanet REsearch instrument (SPHERE) at the Very Large Telescope (VLT), the Gemini Planet Imager (GPI) at the Gemini South Observatory, or the Subaru Coronagraphic Extreme Adaptive Optics (SCExAO) at the Subaru Telescope \citep{Beuzit2019, Macintosh2015, Jovanovic2015}. The coming years will witness the development of, among others, METIS \citep{Kenworthy2016, Brandl2018}, HARMONI \citep{Carlotti2019, Thatte2021, Houlle2021}, and MICADO \citep{Clenet2019, Davies2021} at the Extremely Large Telescope (ELT) and the upgrades of existing instruments like SPHERE+ (VLT, 2025) \citep{Boccaletti2020} and GPI2.0 \citep{Chilcote2020}.

\begin{figure}
   \begin{center}
   \begin{tabular}{c}
   \includegraphics[width=8.5cm]{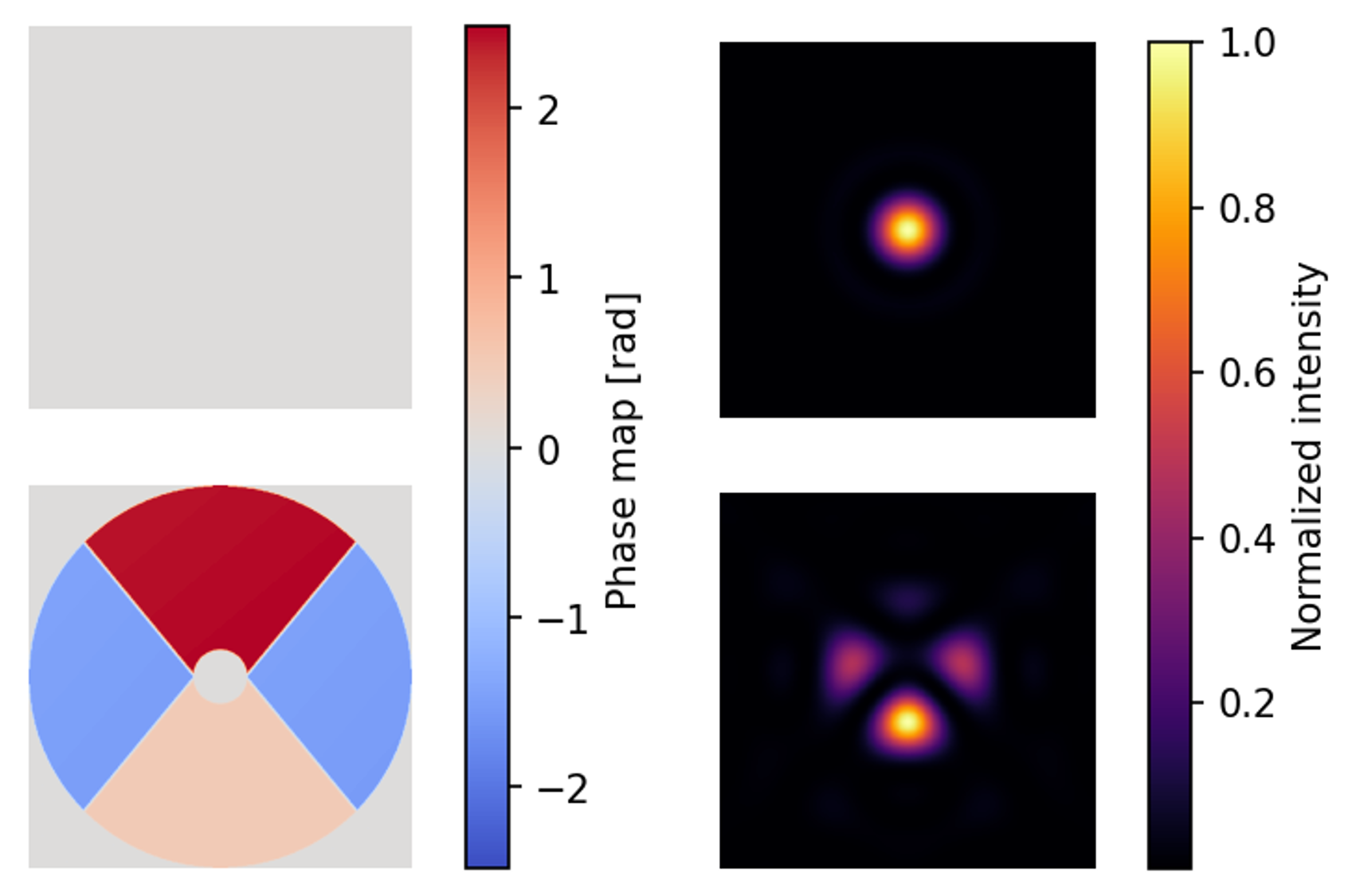}
   \end{tabular}
   \end{center}
   \caption[Fig0_VLTMickeyMouse] 
   { \label{fig:Fig0_VLTMickeyMouse} 
Illustration of the so-called "Mickey Mouse effect" at a VLT-like configuration: (left) aberration maps applied on the pupil, (right) associated PSFs in linear scale (top) without aberration, (bottom) with 415 nm RMS of piston error on the petals.} 
   \end{figure}

However, instruments already in operation, such as VLT/SPHERE or Subaru telescope/SCExAO have revealed themselves sensitive to a specific aberration, called the low-wind effect (LWE), a phenomenon that happens for high emissivity spider telescopes when the wind speed at the level of the telescope aperture is lower than a few m/s \citep{Sauvage2015, Milli2018, Vievard2019, Holzlohner2021}, which becomes dominant under good observing conditions (low seeing). With LWE, the phase is fragmented into differential piston, tip, and tilt on the pupil petals delimited by the spiders that cannot be corrected by traditional Adaptive Optics (AO) systems \citep{N'Diaye2018}. It therefore strongly degrades AO-corrected images and impacts coronagraph performance. 
On non-coronagraphic images, secondary lobes (from one to the number of pupil petals) appear around the location of the first Airy ring, forming the so-called "Mickey Mouse effect" illustrated in Fig. \ref{fig:Fig0_VLTMickeyMouse} \citep{Sauvage2015, Lamb2017, Milli2017}. These secondary lobes act like off-axis sources on coronagraphic systems, which can create a severe starlight leakage through the focal-plane mask. The contrast is then degraded at short separations: on SPHERE, where the LWE is one of the main limitations with a phase error of about 200nm RMS \citep{Sauvage2015, Sauvage2016, Cantalloube2019}, the contrast is degraded by a factor of about 50 at 0.1'', and the spider diffraction patterns usually masked by the Lyot stop remain visible up to several arcseconds. In addition, spider diffraction spikes are also visible out of the AO-correction area. In 2017, a specific coating with a low thermal emissivity in the mid-infrared was applied on the spiders of the Unit Telescope (UT) 3 of the VLT, but despite a drastic decrease in the aberration amplitude, the LWE still impacts the coronagraphic performance necessitating the implementation of additional mitigation strategies \citep{Wilby2018, Vievard2019, Bos2020, Pourre2021}.

The LWE is also expected on upcoming giant telescopes like the ELT ($51$ cm thick spider) and the Thirty Meter Telescope (TMT, $225$ mm thick spider) \citep{Holzlohner2021}. In addition, other island phenomena are expected, in particular the so-called petaling effect. The latter happens when the width of the spiders is larger than the Fried parameter $r_0$ of the atmospheric turbulence at the sensing wavelength. The AO wavefront sensor mistakenly reconstructs discontinuities along the spiders and in particular petal-level pistons \citep{Schwartz2018, Bertrou-Cantou2022}. These two aberrations, i.e. low-wind effect and petaling, will be the purpose of this paper.

In this paper, we propose to adapt the pupil apodization according to its segmentation by the spiders to make it robust to differential piston, tip, and tilt errors between the pupil petals. This concept derives from the Redundant Apodized Pupil (RAP) coronagraphs, developed and tested in \cite{Leboulleux2022}, albeit for primary mirror segmentation-due errors. In section \ref{s:Low-wind effect and petaling effect}, we summarize the island effects expected on the ELT. In section \ref{s:Reminders}, we present a short review of the RAP design methodology and concept and extend and apply it in section \ref{s:ELT} to the ELT pupil to propose alternative apodizers robust to LWE and post-AO petaling. Another short application of this concept to the TMT aperture is also proposed in Appendix \ref{s:Appendix}.

\section{Low-wind effect and petaling effect}
\label{s:Low-wind effect and petaling effect}

Island effects can be discriminated into different phenomena depending on their source: 1) the low-wind effect, which is due to a differential thermal cooling on each side of the spiders and generates petal-level pistons, tips, and tilts, with a typical lifetime $\sim 2-3$ seconds \citep{Milli2018}, and 2) post-AO petaling, due to the inability of the wavefront sensor to reconstruct wavefront continuities when the spiders are much thicker than the atmosphere spatial coherence length $r_0$, which generates petal-level pistons with a typical lifetime of one second \citep{Bertrou-Cantou2020} (see also section \ref{s:Petaling} for petaling-induced piston sequences). These lifetimes, if different than the lifetimes of other aberrations \citep[for instance non common path aberrations, e.g.][]{Vigan2022}, can increase the complexity of the wavefront reconstruction and correction.

The ELT, with its 51 cm-thick spiders, might be subject to these island effects. While the low-wind effect has so far mainly been studied on existing telescopes like VLT or Subaru, studies are also ongoing for the ELT \citep{Holzlohner2021} and in this paper, we assume that LWE has the same properties as those observed on SPHERE, i.e., primarily consisting of differential pistons, tips, and tilts. On the other hand, petaling has mainly been studied in simulation and appears as a strong limitation for HARMONI and MICADO that both use visible pyramid wavefront sensors \citep{Schwartz2018, Bertrou-Cantou2022}.

\begin{figure*}
   \begin{center}
   \begin{tabular}{c}
   \includegraphics[width=13cm]{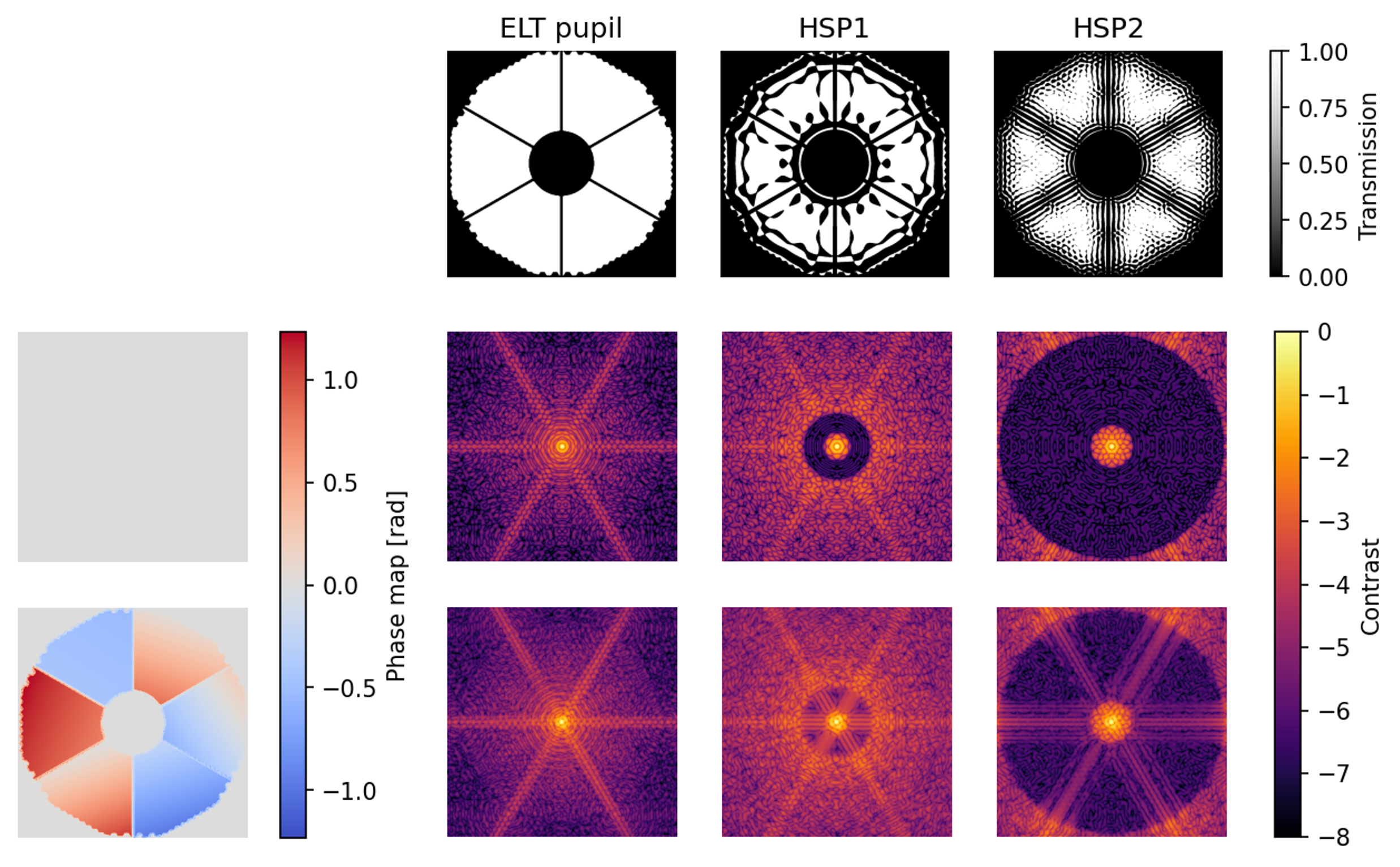}
   \end{tabular}
   \end{center}
   \caption[Fig1_Definitions] 
   { \label{fig:Fig1_Definitions} 
Low-wind effect on the ELT: (top) three pupil configurations with the full ELT pupil and the Shaped pupils HSP1 and HSP2 of HARMONI/HCM, and their associated PSFs in logarithmic scale (center) without low-wind effect, (bottom) with the low-wind effect phase map of the left column. The phase map is composed of petal-level pistons, tips, and tilts with average values and variances equivalent to the ones of VLT/SPHERE before the spider coating (here $\sim 110$ nm RMS of piston and $\sim 180$ nm RMS of tip and tilt at $1.6$ $\mu$m).} 
   \end{figure*}

For both direct and coronagraphic images, these island effects deteriorate the contrast performance. Fig. \ref{fig:Fig1_Definitions} shows the PSFs without and with low-wind effect, at a level equivalent to the one observed on SPHERE before the 2017 spider coating (here, $\sim 110$ nm RMS of piston and $\sim 180$ nm RMS of tip and tilt at $1.6$ $\mu$m) \citep{Pourre2021}, and for three different configurations: without coronagraph, with the HSP1 and HSP2 Shaped Pupils (SP), both designed for the HARMONI high-contrast module (HCM) \citep{Carlotti2018, Henault2018}. Both coronagraphs are capable of reaching a contrast limit of $10^{-6}$ from $5$ to $11\lambda/D$ for HSP1 and from $7.5$ to $40\lambda/D$ for HSP2 (see table \ref{tab:Specs}). These aberrations create spider diffraction spikes that deteriorate the AO-correction area, mainly in the high-contrast dark zone, with an impact on the overall contrast, and particularly contrast at small angular separations. 

For detections limited by speckle noise, for instance when combined with Reference star Differential Imaging (RDI, \cite{Lafreniere2009, Soummer2012, Gerard2016}) and Angular Differential Imaging (ADI, \cite{Marois2006}), static and quasi-static aberrations must also be strictly controlled with the adaptive optics loop. Due to the difference in evolution lifetimes between dynamic adaptive optics residuals and non common path aberrations that require an additional wavefront sensor, island effects can limit the detection unless a specific mitigation tool is provided.

Most mitigation strategies rely on an efficient estimation of the island effect phase errors. So far, different wavefront sensors have been tested: the pyramid wavefront sensor in visible light appears in simulation blind to such errors \citep{Bertrou-Cantou2022}, while the Shack-Hartmann sensor cannot measure petal-level piston errors \citep{Sauvage2016, Pourre2021}, even as promising work is ongoing for the estimation of tip-tilt errors in the context of SPHERE and GRAVITY+ \citep{Pourre2021}. Zernike sensors (ZELDA, which stands for Zernike sensor for Extremely Low-level Differential Aberrations) have successfully been tested on SPHERE to detect and calibrate low-wind effect \citep{Sauvage2015,N'Diaye2016}. Last but not least, several focal-plane wavefront sensors have been tested in simulation, at Subaru, Keck, and the VLT. Phase diversity algorithms, from classical methods \citep{Lamb2016, Lamb2017} to more complex extensions like the Fast \& Furious algorithm \citep{Korkiakoski2014, Wilby2016, Wilby2018} have proven to be efficient in estimating low-wind effect phase errors on SPHERE. On SCExAO, low-wind effect has already been studied with the Asymmetric Pupil Fourier focal-plane wavefront sensor \citep{Martinache2013, N'Diaye2018, Vievard2019}. Still on SCExAO, the phase diversity with the linearized analytic phase diversity technique and once again the Fast \& Furious algorithm \citep{Vievard2019} have also been implemented. The last mitigation technique does not include any wavefront sensor and consists of the application of a coating with low thermal emissivity in the mid-infrared on the spiders to physically reduce thermal gradients \citep{Milli2018}. 

For petaling compensation, except for an infrared pyramid sensor for the METIS AO system \citep{Hippler2019}, \cite{Schwartz2018} proposes an update of the AO correction loop called the edge actuator coupling that consists of considering actuators of each side of the spider as one single actuator to force the system to limit differential pistons between petals. \cite{Bertrou-Cantou2020, Bertrou-Cantou2022} also proposes an update of the reconstruction mode basis to force the phase continuity between petals. Moreover, \cite{Hutterer2018} proposes a split approach with a separated reconstruction of petal-level pistons and of other modes using Bayesian statistics.

This paper investigates a different mitigation strategy, with an adaptation of the RAP design proposed in \cite{Leboulleux2022}. All the island effects mentioned earlier will later be considered as differential pistons between petals, with tips and tilts in the low-wind effect case, and with a quite slow dynamic evolution (one to a few seconds). 

\section{Review of Redundant Apodized Pupils}
\label{s:Reminders}

For segmented telescopes equipped with high-contrast imagers, the Pair-based Analytical model for Segmented Telescopes Imaging from Space \citep[PASTIS][]{Leboulleux2018, pastis, Laginja2021} provides an expression for the intensity in the coronagraphic dark hole as a function of the segment-level phase errors. For a symmetric dark zone with amplitude aberrations and phase aberrations after the focal plane mask of the coronagraph neglected, then the intensity in the dark zone $I$ can be expressed as follows:
\begin{equation}
\label{eq:PASTIS}
    I(\mathbf{u}) = \left \Vert \widehat{Z}(\mathbf{u}) \right \Vert ^2 \times \sum_{k_1=1}^{n_{seg}} \sum_{k_2=1}^{n_{seg}} c_{k_1,l} a_{k_1,l} c_{k_2,l} a_{k_2,l} \cos((\mathbf{r_{k_2}} - \mathbf{r_{k_1}} ). \mathbf{u})
\end{equation}
where $n_{seg}$ is the number of segments in the primary mirror, $\mathbf{u}$ is the position vector on the detector, $\widehat{Z}$ is the Fourier transform of the Zernike polynomial $Z$ considered as segment-level aberrations, $(c_k)_{k\in \lbrack 1,n_{seg} \rbrack}$ are calibration coefficients including the effect of the coronagraph, $(a_k)_{k\in \lbrack 1,n_{seg} \rbrack}$ are the local Zernike coefficients on the segments, and $(r_k)_{k\in \lbrack 1,n_{seg} \rbrack}$ correspond to the positions vectors from the center of the pupil to the centers of the segments.

This equation corresponds to a low-order envelope $\left \Vert \widehat{Z}(\mathbf{u}) \right \Vert ^2$ multiplied by interference fringes between all pairs of segments. For piston-like aberrations, this low-order envelope, also called the halo in \cite{Yaitskova2002, Yaitskova2003} corresponds to the Point Spread Function (PSF) of the segment. The full pupil PSF can then be seen as the segment PSF modulated by the interference fringes between all pairs of segments. The image intensity at each point is proportional to this segment PSF (or low-order envelope).

The RAP concept \citep{Leboulleux2022} consists of apodizing each segment to dig a dark hole in the low-order envelope. Since the image corresponding to the full segment pupil is proportional to the segment PSF, it also deepens the contrast in the segmented pupil coronagraphic PSF. It implies that for a given target contrast in the dark zone $I$, the piston-level error amplitudes $(a_k)_{k\in \lbrack 1,n_{seg} \rbrack}$ can be larger if the low-order envelope $\left \Vert \widehat{Z}(\mathbf{u}) \right \Vert ^2$ is reduced (see Eq. \ref{eq:PASTIS}), meaning an increased robustness to segment-level aberrations. 

In \cite{Leboulleux2022}, the RAP concept is applied on errors due to the segmentation of the primary-mirror, on a Giant Magellan telescope-like pupil combined with an Apodized Pupil Lyot Coronagraph (APLC) \citep[e.g., ][]{Soummer2009, N'Diaye2015, Zimmerman2016} on one hand and with an Apodizing Phase Plate (APP) \citep[e.g., ][]{Codona2007, Kenworthy2010, Carlotti2013, Por2017} coronagraph on the other hand. RAPs deliver an increased robustness albeit at more modest angular separations than targeted by exo-Earth-imagers. As another advantage, the RAP is a passive component that does not require any additional wavefront sensor or wavefront control tool like a deformable mirror to reduce the constraints on the optical wavefront.

We propose to consider the petals delimited by the secondary mirror spiders as pupil segmentation, low-wind effect as segment-level piston, tip, and tilt errors, and post-AO petaling as segment-level piston errors. With these conditions, the RAP concept can be extended and applied to generate coronagraphs robust to island effects and in the next section, we focus on ELT applications. Its six petals are identical, a configuration fitting with the RAP procedure.

Usually when computing a coronagraph apodization, it is common to reduce the computational burden of the apodization optimization by apodizing only one quarter of the pupil, when it has a two-axis symmetry \citep{Por2020, Carlotti2014}. The full pupil PSF can then be numerically computed from one quarter of the pupil only, and the apodization obtained on one quarter from this optimization process is then unfolded to obtain the full pupil apodization. The RAP procedure, with an optimization by petal, should not be compared to this numerical trick: it aims at improving the contrast in the low-order envelope instead of the full pupil PSF, and the full pupil PSF is never computed in the apodization optimization process.

\section{Application to the island effect mitigation on the ELT}
\label{s:ELT}

\subsection{Apodizer design}
\label{s:Apodizer design}

\subsubsection{Requirements and methodology}

The objective is to design both amplitude (SP, \cite{Vanderbei2003, Kasdin2003, Carlotti2011}) and phase (APP for Apodizing Phase Plate, \cite{Codona2007, Kenworthy2010, Carlotti2013, Por2017}) apodizers at the scale of one petal and to reproduce them to cover the entire ELT pupil.

The apodization problem consists of minimizing the intensity in the defined dark region by exploring the parameter space of the petal apodization while maximizing the throughput of the final apodizer. As in \cite{Leboulleux2022}, it is formulated in AMPL (A Mathematical Programming Language) \citep{Fourer2002}, and the procedure then calls the Gurobi solver to find the optimal solution \citep{gurobi}.

On the HCM of HARMONI, two designs have been selected: HSP1 and HSP2 (see Fig. \ref{fig:Fig1_Definitions}), whose specifications (IWA for Inner Working Angle, OWA for Outer Working Angle, and contrast) are reminded in Table \ref{tab:Specs} \citep{Carlotti2018}. The requirements for the RAP designs are equivalent to the ones of HSP2, i.e. set at a deepest contrast of $10^{-6}$ from $8$ to $40 \lambda/D$, where $\lambda$ is the wavelength (later on set at $1.6 \mu$m) and $D$ the pupil diameter ($39$m) (see also Table \ref{tab:Specs}).

\begin{table}
    \caption{Specifications for the four cases, HSP1 and HSP2 being specifically designed by A. Carlotti for the HCM, and the two RAP cases developed later on.}
    \label{tab:Specs}
    \centering
    \begin{tabular}{l||c|c|c|c}
        & HSP1 & HSP2 & SP RAP & APP RAP \\
        \hline \hline
        IWA & $5 \lambda/D$ & $7.5 \lambda/D$ & $8 \lambda/D$ & $8 \lambda/D$ \\
        \hline
        OWA & $11 \lambda/D$ & $40 \lambda/D$ & $40 \lambda/D$ & $40 \lambda/D$ \\
        \hline
        Contrast & $10^{-6}$ & $10^{-6}$ & $10^{-6}$ & $10^{-6}$ \\
    \end{tabular}
\end{table}

Because there are approximately $2.4$ petals along the ELT pupil diameter, these requirements correspond to a dark zone between $3.5$ and $16.3 \lambda/d$ for the petal low-order envelope, where $d$ is the size of the petal. As in \cite{Leboulleux2022}, accessing smaller IWAs imposes a drastic decrease in planet throughput and pupil transmission; an IWA of $3.5 \lambda/d$ was the best compromise found for this application.

\subsubsection{Amplitude apodizer (SP)}

\begin{figure}
   \begin{center}
   \begin{tabular}{c}
   \includegraphics[width=6.5cm]{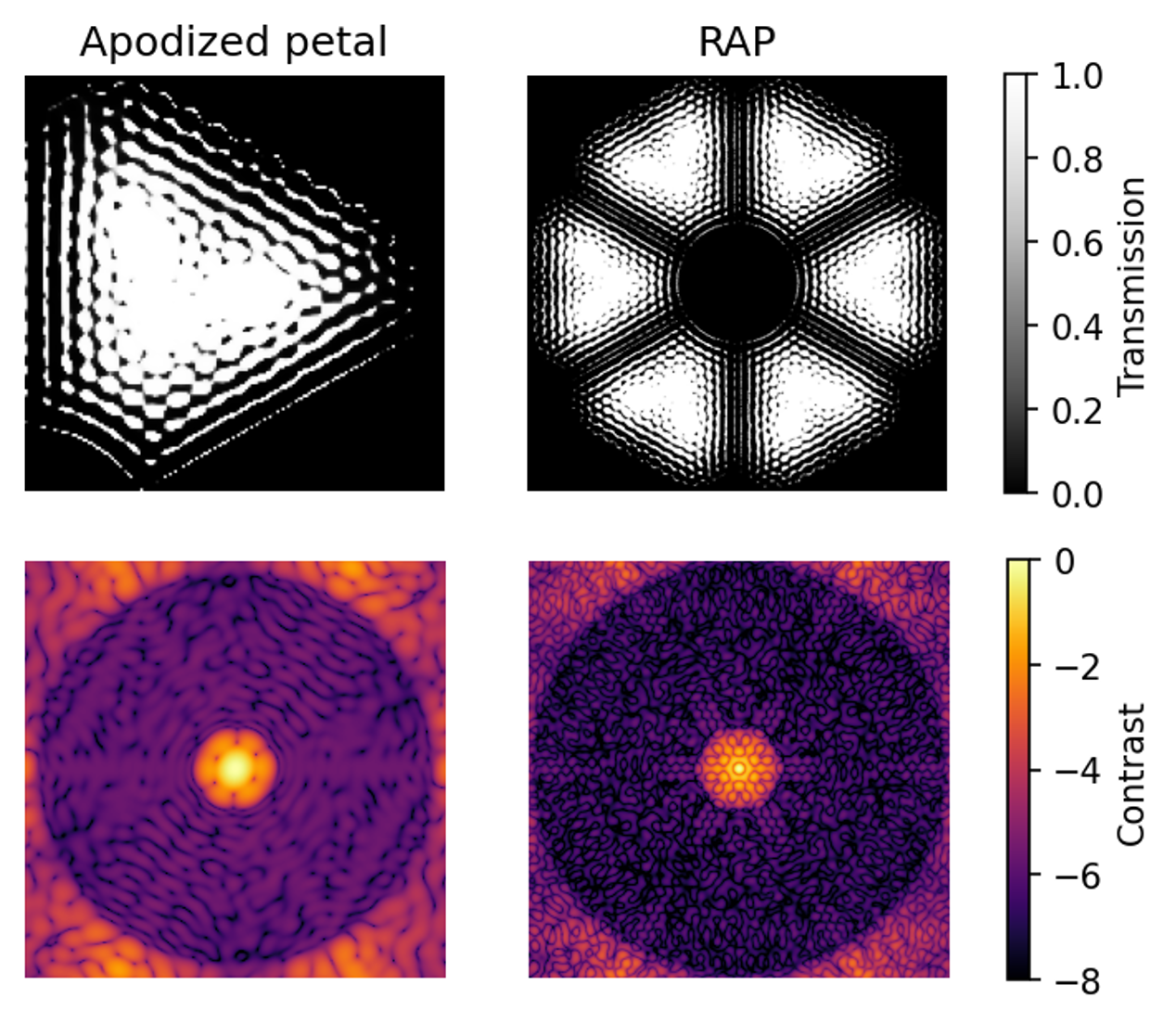}
   \end{tabular}
   \end{center}
   \caption[Fig3_SP] 
   { \label{fig:Fig3_SP} 
Amplitude apodizations: (left) for a single petal, (right) for the full RAP, with (top) the pupil transmissions from 0 (no transmission) to 1 (full transmission), aiming for a contrast of $10^{-6}$ between $8$ and $40 \lambda/D$ and (bottom) associated PSFs in logarithmic scale.} 
   \end{figure}

\begin{figure}
   \begin{center}
   \begin{tabular}{c}
   \includegraphics[width=8.5cm]{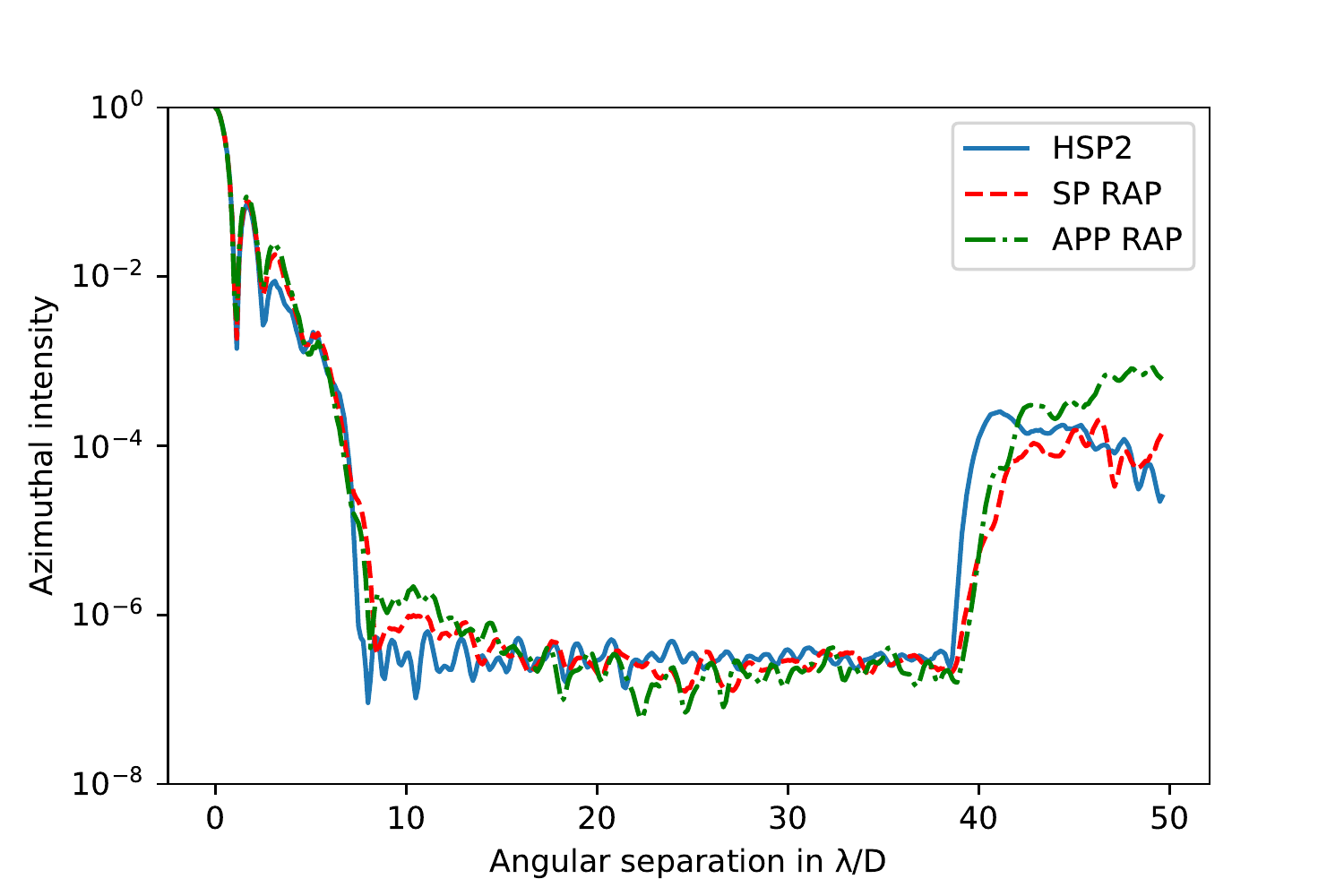}
   \end{tabular}
   \end{center}
   \caption[Fig4.5_AzimuthalCurves] 
   { \label{fig:Fig4.5_AzimuthalCurves} 
Coronagraphic intensity profiles, normalized to the intensity peak, of one of the two Shaped Pupils designed for the HARMONI configuration (HSP2 in blue), and the two RAPs presented in this paper: (red) the amplitude RAP and (green) the phase RAP. The three designs aim for $10^{-6}$ contrasts in their dark zones (see Table \ref{tab:Specs}).}
   \end{figure}

The petal amplitude apodization issued from the AMPL procedure is visible in Fig. \ref{fig:Fig3_SP} (left), as well as its PSF, which is also the low-order envelope of the full RAP. This later is obtained by rotating the petal apodizer around the pupil center by angular steps of $60^{\circ}$. The output design and its PSF are also shown in Fig. \ref{fig:Fig3_SP} (right). We observe that the petal apodizer PSF does not require the contrast constraint to be set at $10^{-6}$, $10^{-5.25}$ in the petal PSF dark zone being sufficient to obtain a $10^{-6}$ contrast on the final RAP PSF dark zone. The intensity profile of the designed apodizer is also shown in Fig. \ref{fig:Fig4.5_AzimuthalCurves} (red).

Table \ref{tab:Throughputs} also presents the throughputs $T$ (with respect to the definition of \cite{Ruane2018}) and transmissions $t$ of the final designs, which are lower than the corresponding values for HSP2 and defined as follows:
\begin{equation}
\label{eq:Throughput}
    t = \frac{\int RAP(\vec{x}) d\vec{x}}{\int P(\vec{x}) d\vec{x}} \text{~~and~~}
    T = \frac{\int _{A} I_{c}(\vec{u}) d\vec{u}}{\int _{A} I_{nc}(\vec{u}) d\vec{u}}
\end{equation}
with $RAP$ the RAP function (full apodizer with $0$ if no transmission, $1$ if transmission), $P$ the entrance pupil function, $A$ a $0.7 \lambda/D$ radius circular area centered on the star image, $I_c$ the coronagraphic image and $I_{nc}$ the non-coronagraphic image.

\begin{table}
    \caption{Throughputs and transmissions of the three apodizers.}
    \label{tab:Throughputs}
    \centering
    \begin{tabular}{l||c|c|c}
        & HSP2 & SP RAP & APP RAP \\
        \hline \hline
        Throughput & $32 \%$ & $24 \%$ & $18 \%$ \\
        \hline
        Transmission & $55 \%$ & $48 \%$ & $100 \%$ \\
    \end{tabular}
\end{table}

\subsubsection{Phase apodizer (APP)}

\begin{figure}
   \begin{center}
   \begin{tabular}{c}
   \includegraphics[width=6.5cm]{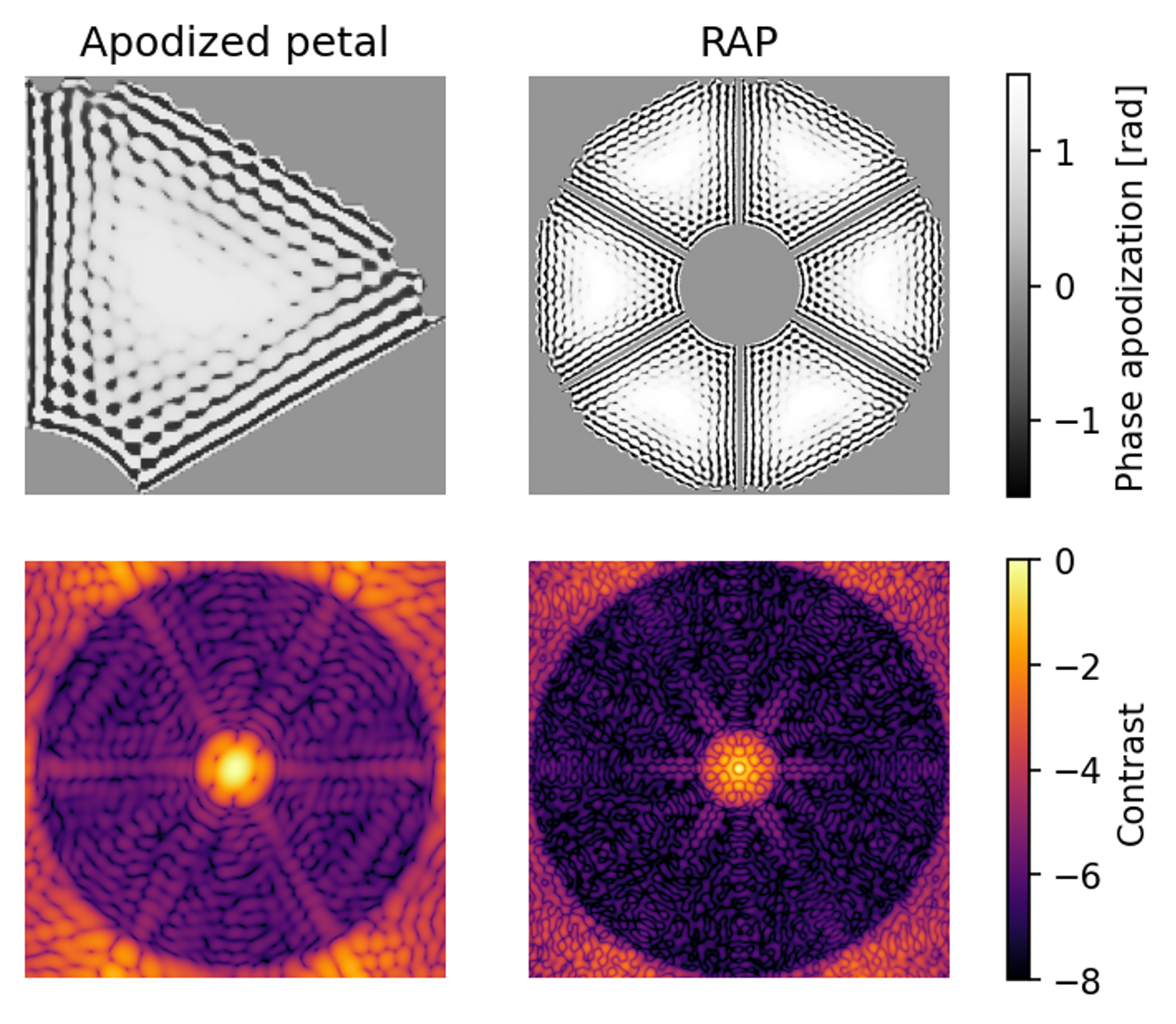}
   \end{tabular}
   \end{center}
   \caption[Fig4_APP] 
   { \label{fig:Fig4_APP} 
Phase apodizations: (left) for a single petal, (right) for the full RAP, with (top) the pupil phases in radians, aiming for a contrast of $10^{-6}$ between $8$ and $40 \lambda/D$ and (bottom) associated PSFs in logarithmic scale.} 
   \end{figure}

In Fig. \ref{fig:Fig4_APP}, one can see the petal phase apodization computed by the AMPL code, in addition to its associated PSF. Once again, the RAP is designed from 6 rotations of this petal-level apodizer, also visible in the figure with its PSF. The final design reaches the target contrast of $10^{-6}$ between $8$ and $40 \lambda/D$, with throughput and transmission values provided in Table \ref{tab:Throughputs}. The $100\%$ transmission in this case does not reflect the loss of planetary photons despite no amplitude apodization and the transmission value should be ignored. The throughput brings a more accurate information about the ratio of planetary photon loss. Patterns close to the IWA are visible on the RAP PSF, with a local contrast slightly above $10^{-6}$; similar effects have been obtained for other designs with different specifications. The intensity profile of the output is also shown on Fig. \ref{fig:Fig4.5_AzimuthalCurves} (green), with the slight deterioration of contrast due to the patterns close to the IWA visible around $8-11 \lambda/D$. These residuals are probably due to the diffraction by the edge of the petal and are already present in the single petal image. They are also visible in all alternative designs conducted for this study with petal-level APP. 

\subsection{Robustness to island effects}
\label{s:Robustness to low-wind effect}

Since the RAPs are optimized to be robust to piston-like errors between petals, we separately examine their robustness to piston and tip/tilt errors.

\subsubsection{Petal-level piston errors}

\begin{figure*}
   \begin{center}
   \begin{tabular}{c}
   \includegraphics[width=13cm]{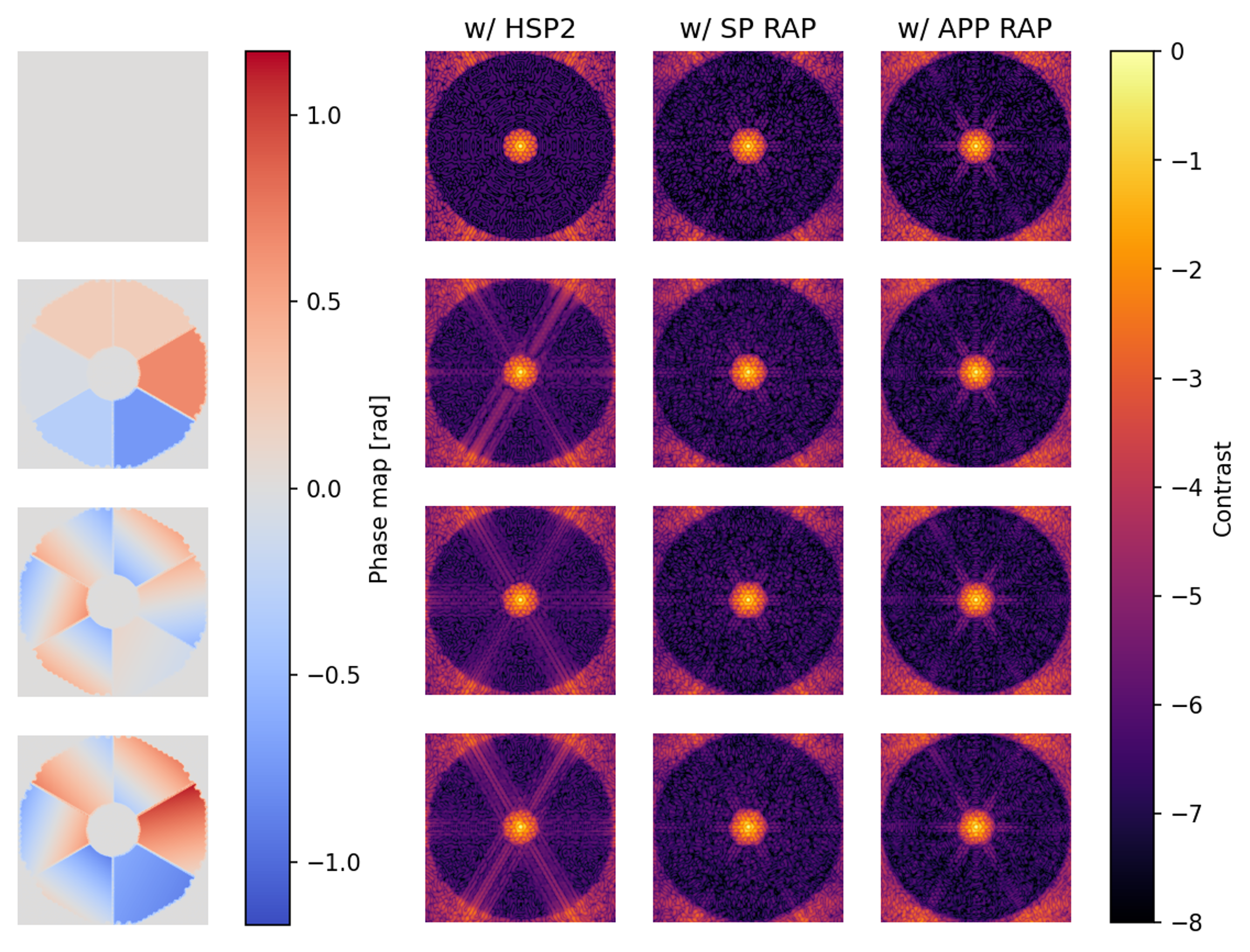}
   \end{tabular}
   \end{center}
   \caption[Fig5_ErrorBudget] 
   { \label{fig:Fig5_ErrorBudget} 
Robustness to low-wind effect errors between petals. (left) phase maps simulated in the pupil plane in radians, (right) associated PSFs in logarithmic scale with the three designs. Four phases are considered: (top) no error, (2nd line) piston only-like errors, at the level of what has been observed on SPHERE before coating, (3rd line) tip-tilt only-like errors, also at the SPHERE level, (bottom) combined piston-tip-tilt errors from above lines.} 
   \end{figure*}

Figure \ref{fig:Fig5_ErrorBudget} (top two lines) shows the PSFs without and with pistons between the pupil petals for the three different apodizers considered here. The amplitudes of the pistons have been chosen with the same statistics (average and variance) as was observed on the SPHERE instrument before the 2017 spider coating. The PSFs from the two RAPs do not appear to be impacted, while strong leakages appear on the PSF behind HSP2. 

\begin{figure}
   \begin{center}
   \begin{tabular}{c}
   \includegraphics[width=8.5cm]{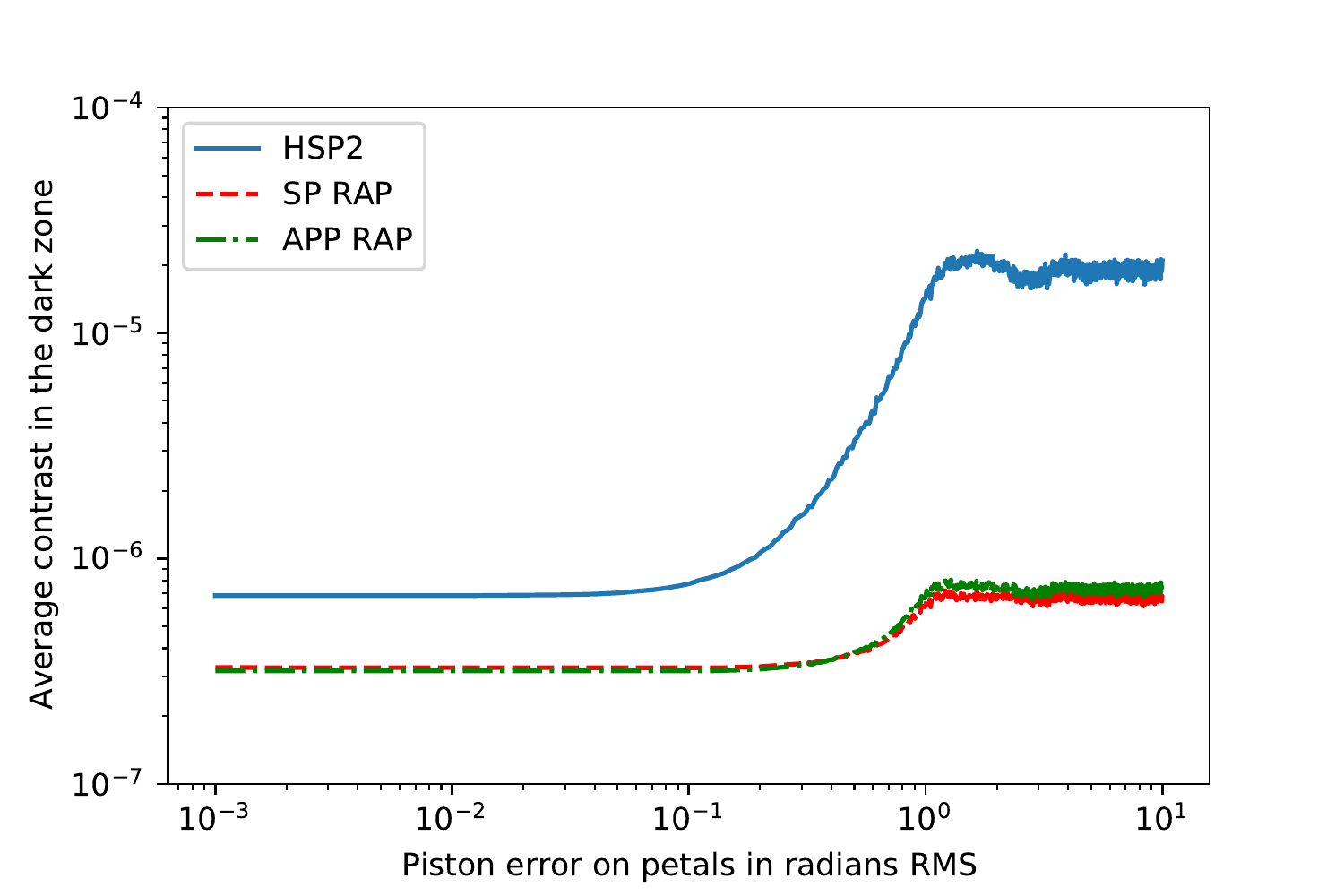}
   \end{tabular}
   \end{center}
   \caption[Fig6_HockeyCross] 
   { \label{fig:Fig6_HockeyCross} 
Evolution of the contrast with the piston-like low-wind effect errors between petals with (blue) the HSP2 design, (red) the amplitude RAP design, (green) the phase RAP design. For each amplitude error considered here, 100 random phase maps are simulated and propagated through the coronagraph, and the average of the 100 contrasts is plotted here.} 
   \end{figure}

More generally, one can compute the average contrast in the dark zone as a function of the amplitude of the aberrations for a large range of aberrations. In Fig. \ref{fig:Fig6_HockeyCross}, amplitudes of aberrations between 1 milliradian (mrad) and 10 radians are considered, and for each amplitude, 100 random phase maps are simulated and propagated through the coronagraph. For each error amplitude, the average of the 100 mean contrasts is computed and appears on this curve. All apodizers have been designed to access a maximum contrast in their dark zone of $10^{-6}$ while these curves indicate the average contrast in the dark zone, which accounts for the values of the plots below $10^{-6}$ for small wavefront error amplitudes. It is noticeable that the two RAP designs show a quite stable performance that remain below $10^{-6}$ despite the increasing aberrations and stabilize around $7\times 10^{-7}$ for aberrations up to $1$ rad RMS, while the HSP2 performance starts to be impacted around $200$ mrad RMS. In particular, the deterioration of contrast at large amplitude aberrations reflects the loss of coherence in the planet image (above $1$ rad).

The PSF deterioration by petal-level piston errors is localized along the axes perpendicular to the spider shadows and more impacted at small angular separations than further from the optical axis. This localized effect is not fully captured by the metric chosen in Fig. \ref{fig:Fig6_HockeyCross}, which is computed over the entire dark zone.

\subsubsection{Petal-level tip-tilt errors}

In Fig. \ref{fig:Fig5_ErrorBudget} (third line), PSFs in presence of only petal-level tip/tilt aberrations are also presented for the three apodizers considered in this study. Like in the previous section, the tip-tilt amplitudes have been randomly selected with average and variances equivalent to the values recorded on SPHERE before the 2017 spider coating \citep{Pourre2021}. In the HSP2 case, petal tips and tilt appear to have a similar effect to the petal pistons, i.e. starlight leakages along the spider diffraction spikes. In the RAP cases though, these spikes are milder.

\begin{figure}
   \begin{center}
   \begin{tabular}{c}
   \includegraphics[width=8.5cm]{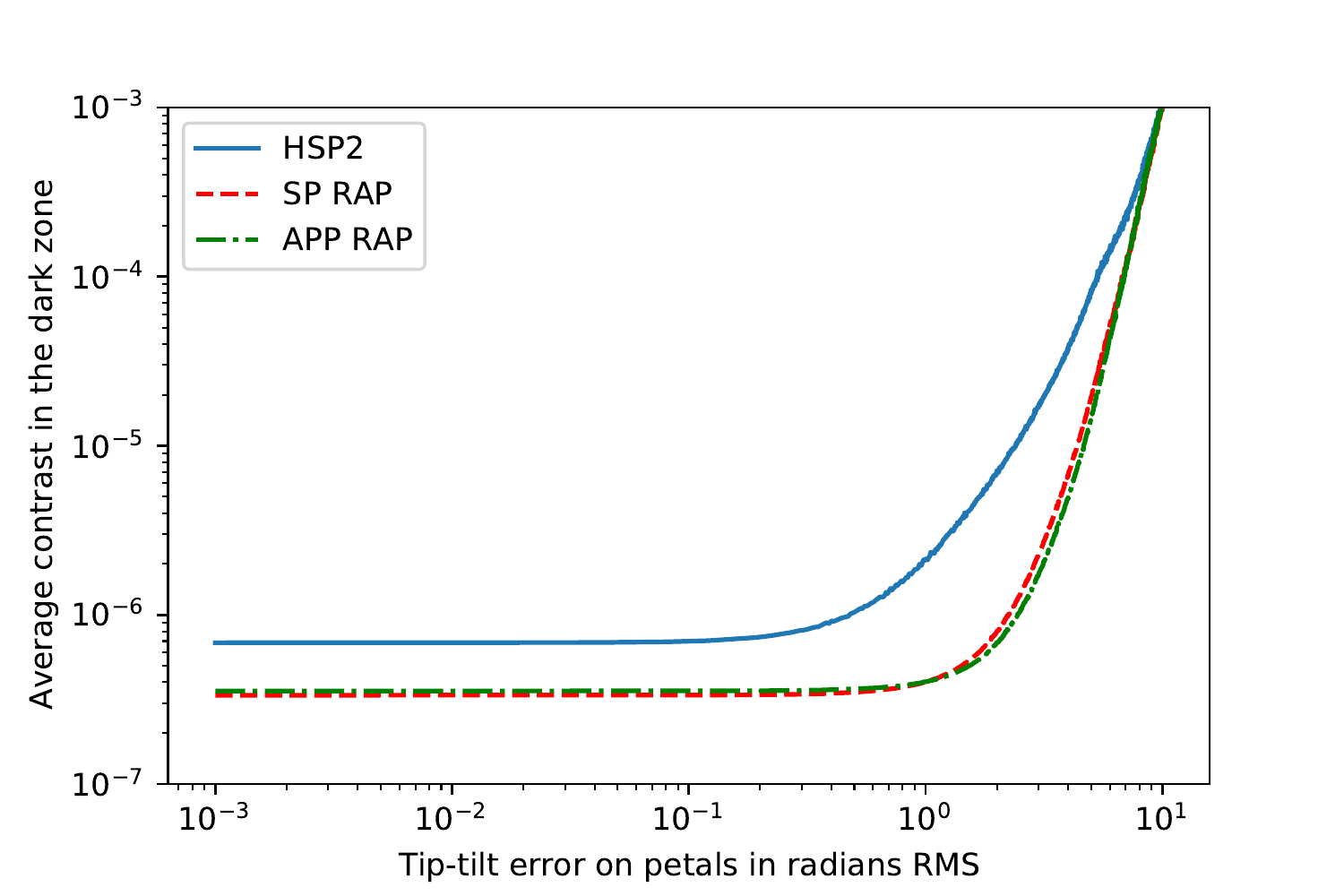}
   \end{tabular}
   \end{center}
   \caption[Fig7_HockeyCrossTT] 
   { \label{fig:Fig7_HockeyCrossTT} 
Evolution of the contrast with the tip-tilt-like low-wind effect errors between petals with (blue) the HSP2 design, (red) the amplitude RAP design, (green) the phase RAP design. For each amplitude error considered here, 100 random phase maps are simulated and propagated through the coronagraph, and the average of the 100 contrasts is plotted here.} 
   \end{figure}

To verify this assumption, we compute the evolution of the average contrast in the dark zone for a large range of petal-level tip-tilt amplitudes. As above, amplitudes from 1mrad to 10 rad are considered, and for each amplitude 100 random phase maps are computed and propagated through the coronagraph, the average value of the 100 associated contrasts being plot on Fig. \ref{fig:Fig7_HockeyCrossTT}. The average contrasts reach $10^{-6}$ at respectively $0.48$, $2.3$, and $2.4$ rad RMS for the HSP2, SP RAP, and APP RAP designs.

The impact of petal-level errors on the coronagraphic PSF is lower for tip-tilt than for piston for the HSP2 design, with local pistons remaining the strongest limitation to the performance.

\subsubsection{Post-OA petaling}
\label{s:Petaling}

Post-AO system petaling is also expected at the ELT, particularly for the MICADO and HARMONI instruments. We propose to study its effect on the coronagraphic performance for different conditions and correction methods, with the AO specifications described in \cite{Bertrou-Cantou2022} (Table 1), i.e. a $500$ Hz loop rate with a visible light ($700$ nm) pyramid wavefront sensor and a $5352$ actuator deformable mirror.

\begin{figure}
   \begin{center}
   \begin{tabular}{c}
   \includegraphics[width=8.5cm]{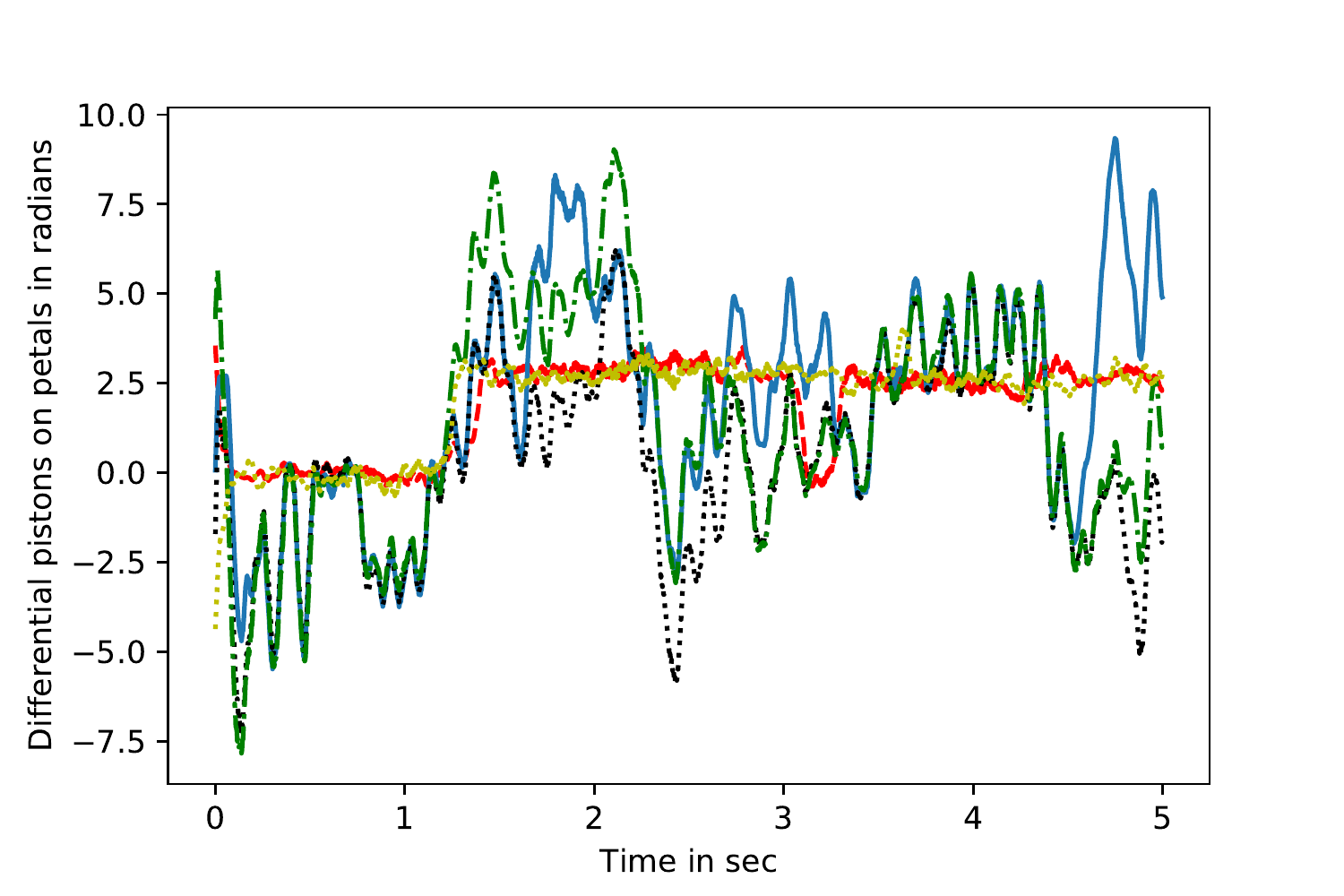} \\
   \includegraphics[width=8.5cm]{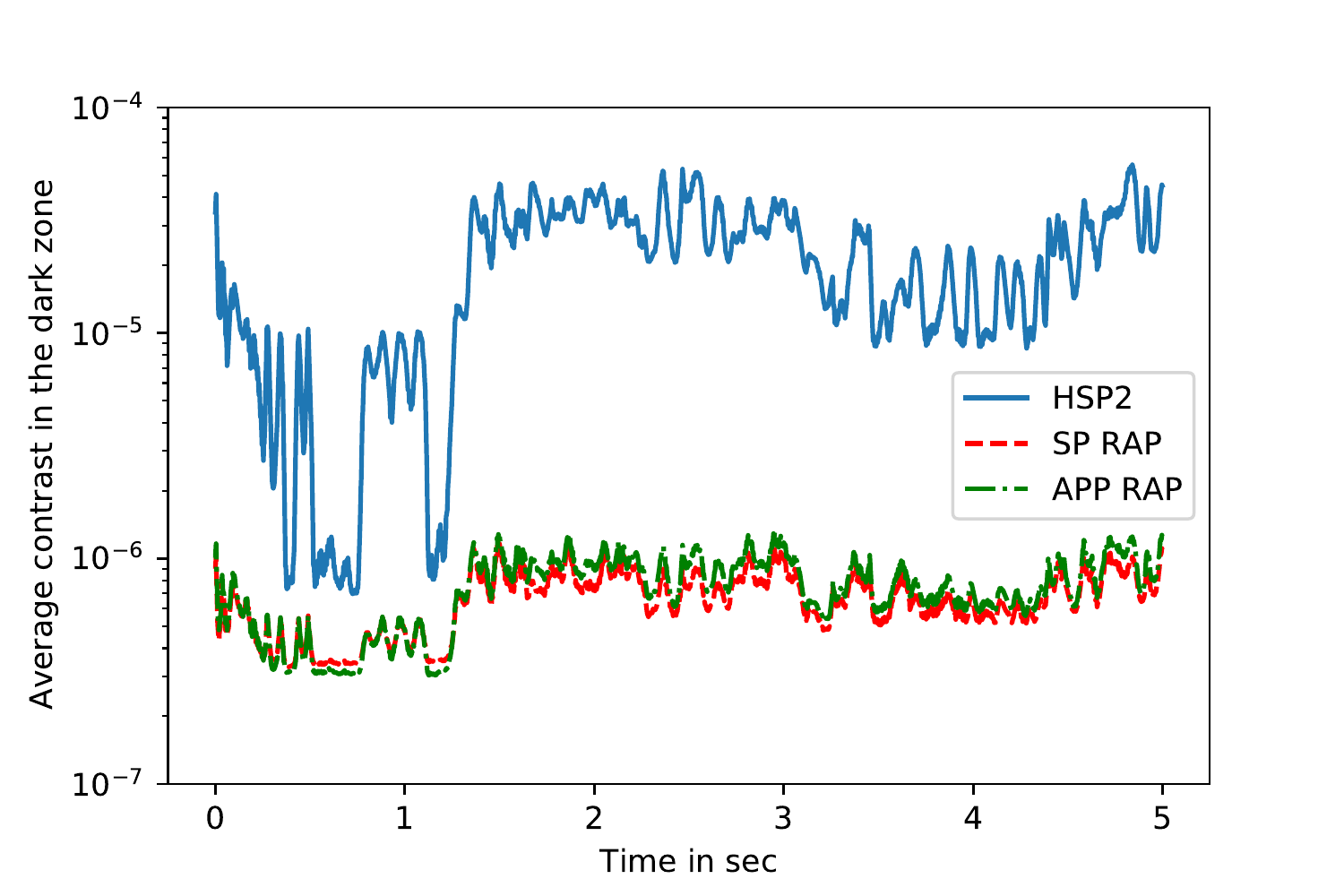}
   \end{tabular}
   \end{center}
   \caption[Fig11_Arielle_Q3Corr1] 
   { \label{fig:Fig11_Arielle_Q3Corr1} 
Impact of the petaling on the coronagraphic performance: (top) differential piston sequence in radians over 5 seconds for 5 petals, the sixth one being the reference, for $r_0 = 12.8$ cm conditions and behind the AO system, (bottom) average contrasts of the three coronagraphs considered in this paper.} 
   \end{figure}

The first considered conditions correspond to a Fried parameter $r_0 = 12.8$ cm at $500$ nm (Q3 conditions as defined by the European Southern Observatory, ESO). The AO system generates differential pistons between segments, plotted in Fig. \ref{fig:Fig11_Arielle_Q3Corr1} (top). The data is issued from the COMputing Platform for Adaptive opticS System (COMPASS) end-to-end simulation software \citep{Ferreira2018, Ferreira2018b, Conan2005}. Fig. \ref{fig:Fig11_Arielle_Q3Corr1} (bottom) also presents the average contrasts in their dark zones behind the different coronagraphs considered in this paper.

\begin{figure}
   \begin{center}
   \begin{tabular}{c}
   \includegraphics[width=8.5cm]{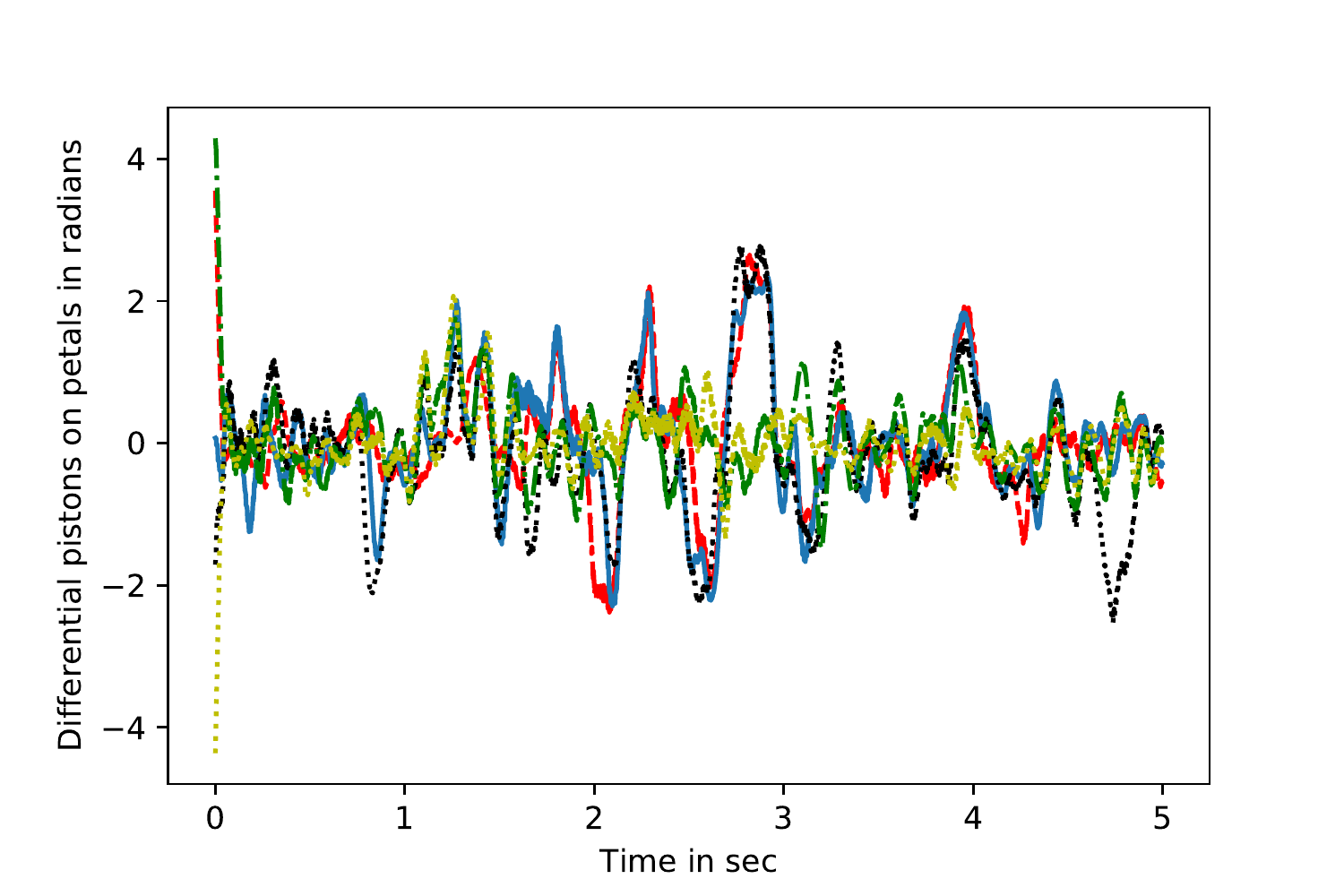} \\
   \includegraphics[width=8.5cm]{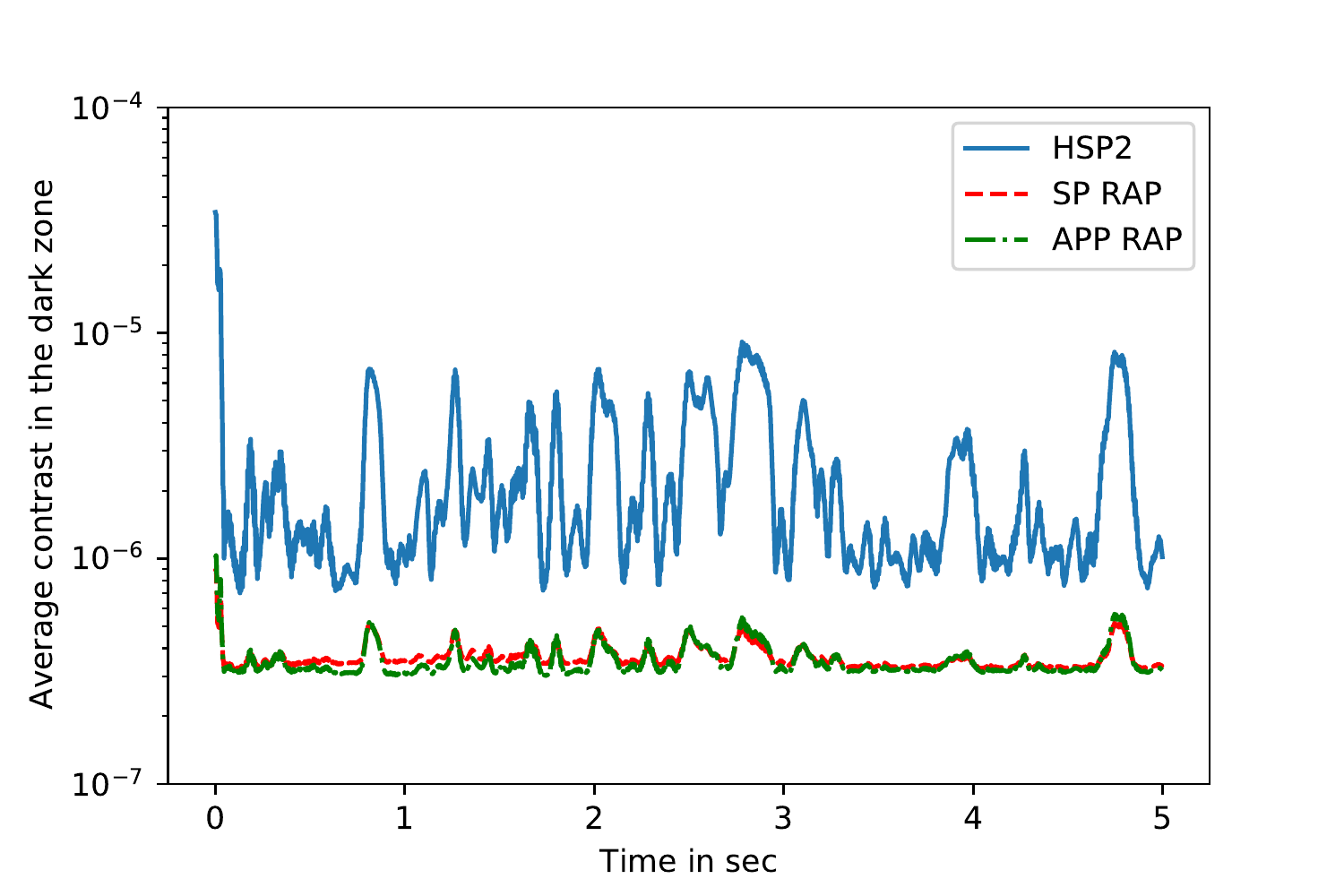}
   \end{tabular}
   \end{center}
   \caption[Fig11_Arielle_Q3Corr2] 
   { \label{fig:Fig11_Arielle_Q3Corr2} 
Impact of the petaling on the coronagraphic performance: (top) differential piston sequence in radians over 5 seconds, for 5 petals, the sixth one being the reference, for $r_0 = 12.8$ cm conditions, behind the AO system and combined with continuity hypotheses between petals, (bottom) average contrasts of the three coronagraphs considered in this paper.} 
   \end{figure}

However, compensation methods are planned to correct for these differential pistons. Fig. \ref{fig:Fig11_Arielle_Q3Corr2} (top) indicates the differential pistons with the same turbulence sequence as in Fig. \ref{fig:Fig11_Arielle_Q3Corr1} but the AO system being coupled with continuity hypotheses between petals \citep{Bertrou-Cantou2020}. Fig. \ref{fig:Fig11_Arielle_Q3Corr2} (bottom) shows the average dark zone contrasts behind the three apodizers.

\begin{figure}
   \begin{center}
   \begin{tabular}{c}
   \includegraphics[width=8.5cm]{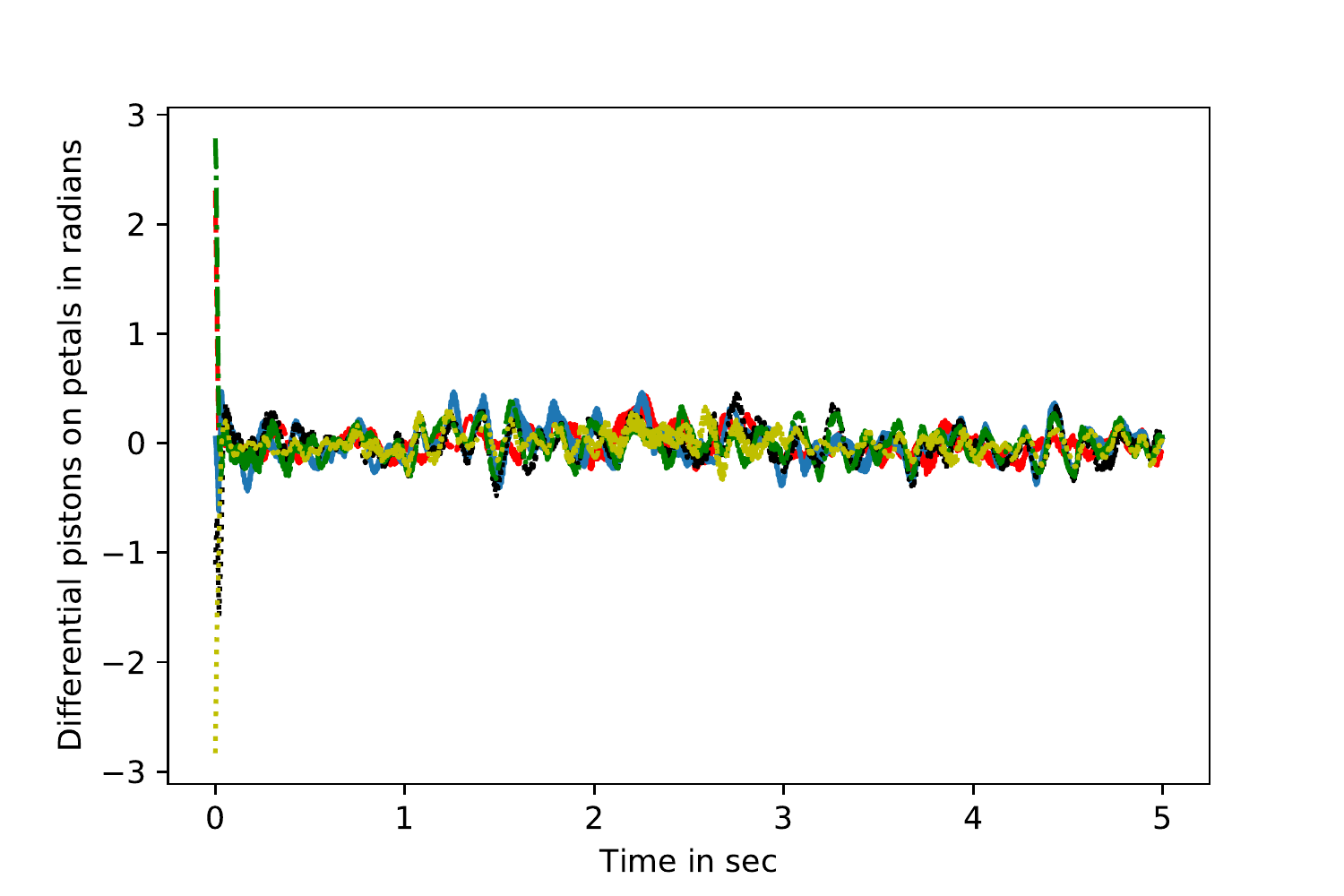} \\
   \includegraphics[width=8.5cm]{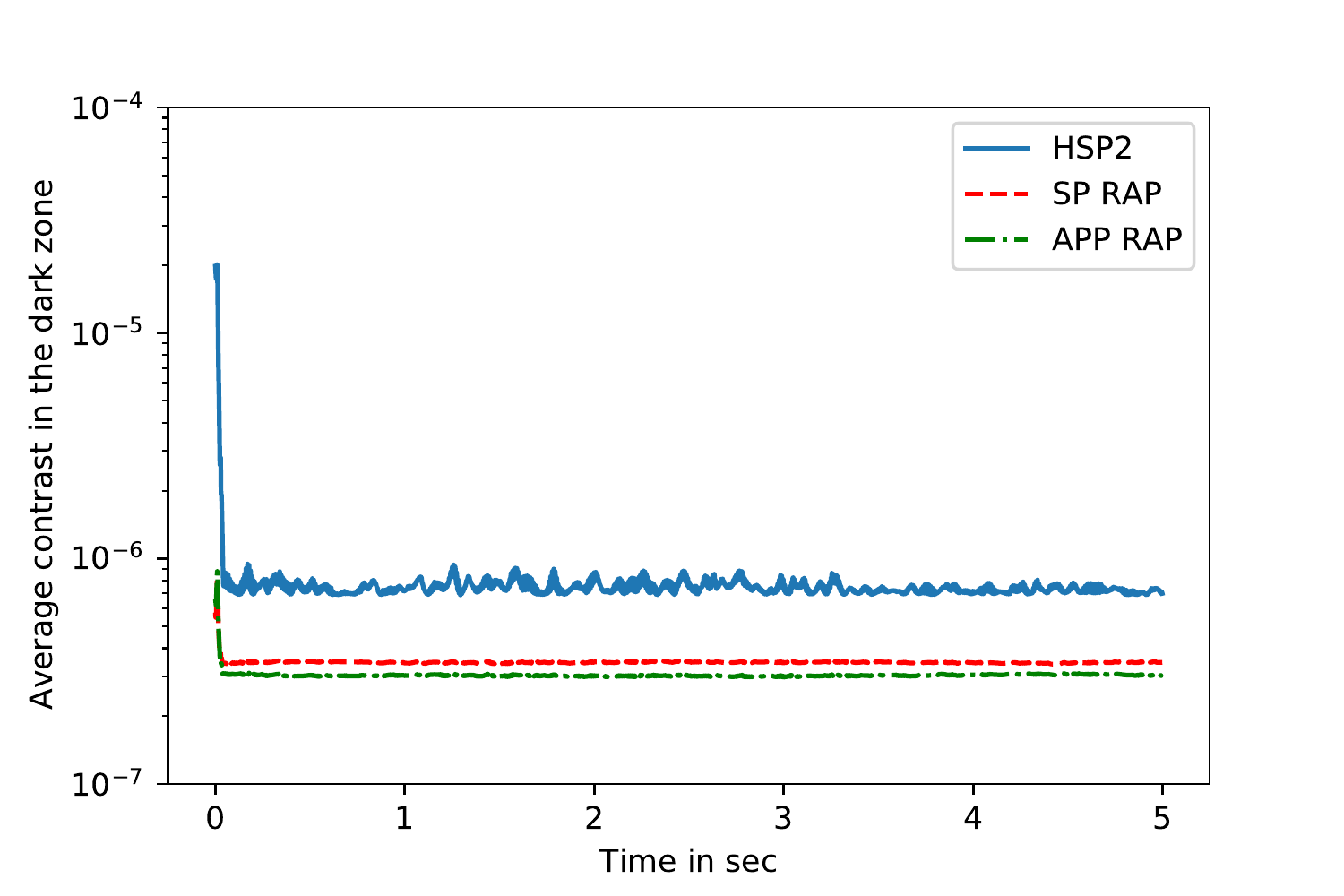}
   \end{tabular}
   \end{center}
   \caption[Fig11_Arielle_Q1Corr2] 
   { \label{fig:Fig11_Arielle_Q1Corr2} 
Impact of the petaling on the coronagraphic performance: (top) differential piston sequence in radians over 5 seconds, for 5 petals, the sixth one being the reference, for $r_0 = 21.5$ cm conditions, behind the AO system and combined with continuity hypotheses between petals, (bottom) average contrasts of the three coronagraphs considered in this paper.} 
   \end{figure}

The RAPs are also tested in milder conditions with a Fried parameter $r_0 = 21.5$ cm at $500$ nm (Q1 conditions as defined by ESO). We consider an AO correction, combined once again with continuity hypotheses between petals. Fig. \ref{fig:Fig11_Arielle_Q1Corr2} presents the differential pistons issued from COMPASS, in addition with the average contrasts in the coronagraphic dark zones.

The RAP designs show a significant robustness with no contrast deterioration to post-AO petaling effects in various atmospheric turbulence conditions. They also have no particular requirement for additional tools such as continuity hypotheses between petals, since their average contrast remains below $10^{-6}$ both without and with these hypotheses.

\subsection{Robustness to other low-order aberrations}
\label{s:Robustness to other low-order aberrations}

In this section, we focus on typical alignment errors, expressed as the first Zernike polynomials, and test the robustness of the three coronagraph designs. 

\begin{figure}
   \begin{center}
   \begin{tabular}{c}
   \includegraphics[width=8.5cm]{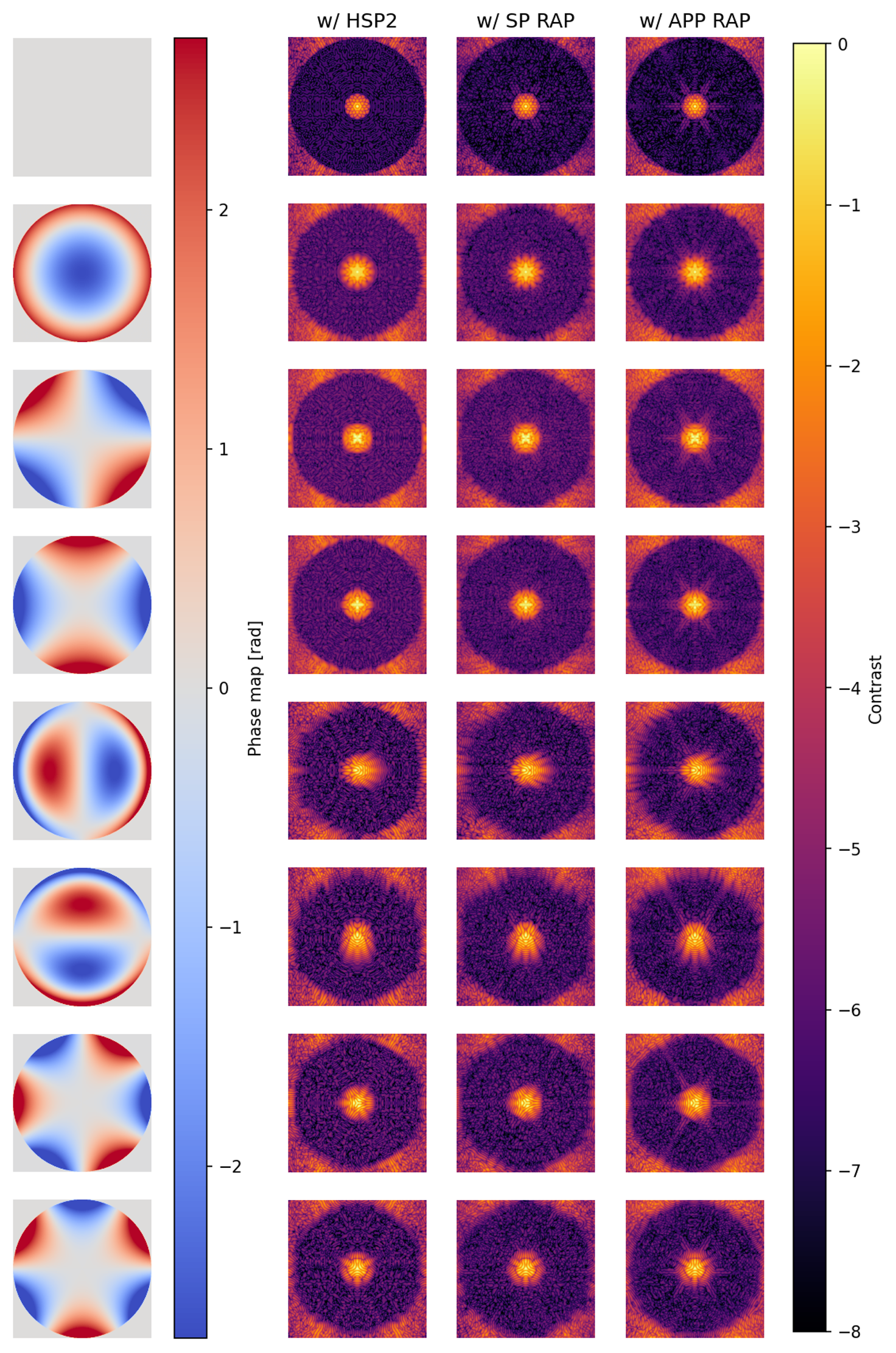}
   \end{tabular}
   \end{center}
   \caption[Fig9_ErrorBudget_Zernike] 
   { \label{fig:Fig9_ErrorBudget_Zernike} 
Impact of low-order aberrations on the coronagraphic PSFs. (right) phase maps in radians, from focus to trefoil, each of them being set at $0.25 \lambda$ RMS except for the first line (no aberration), (right) associated PSFs in logarithmic scale with the three designs, each of them normalized to its intensity peak.} 
   \end{figure}

Fig. \ref{fig:Fig9_ErrorBudget_Zernike} shows the coronagraphic PSFs for the first Zernike polynomials, each of them normalized at $0.25 \lambda$ RMS on a circular support. These aberrations are low order, and mainly modify the shape of the PSF central core in a similar way for the three designs. Since they have an impact close to the optical axis, they correspond to a loss in performance at small angular separations, i.e. an increase of IWA. Despite little to no impact in the dark zone far from the optical axis, the three designs suffer from a similar contrast loss at small angular separations.

In addition, aberrations that impact the coronagraphic PSF at angular separations smaller than the IWA decrease the Strehl ratio by modifying the PSF central core. This Strehl ratio decrease translates into an uniform contrast loss for the entire dark zone, which also appears on Fig. \ref{fig:Fig9_ErrorBudget_Zernike} with an equivalent impact for the three designs.

\section{Conclusions}
\label{s:Conclusions}

This paper follows up a first introduction to the RAP component in \cite{Leboulleux2022}, which was then developed to design coronagraphs robust to primary-mirror segmentation-due errors. In this second paper, we propose an extension of this technique to island effects (low-wind effect and post-AO petaling). 

RAPs applied to island effect are designed by generating an apodizer for one petal only and reproducing it over the other petals to mimic the pupil discontinuity. In this paper, it is applied to the ELT pupil to design both amplitude and phase apodizers (Shaped Pupil and Apodizing Phase Plate) that create dark zones down to $10^{-6}$ between $8$ and $40 \lambda/D$. The resulting apodizers appear extremely robust to petal-level piston errors over the entire range of considered aberrations ($1$ mrad RMS to $10$ rad RMS) and are also highly robust to petal-level tip-tilt errors, with constraints of around $2$ rad RMS for average contrasts in the dark zone of $10^{-6}$.

Similarly to what has been observed in \cite{Leboulleux2022}, RAPs do not allow for IWAs as low as classical apodizers since they are limited by the petal diffraction limit: HSP1 and HSP2 can access angular separations down to $5$ and $7.5 \lambda/D$ respectively, versus $8 \lambda/D$ for both RAPs. For instance, a $7 \lambda/D$ IWA translates into an important loss of throughput compared to a $8 \lambda/D$ IWA, from $24 \%$ to $11 \%$ in the SP RAP case. However, two-step (possibly optimized simultaneously) strategies can be implemented:

\quad \textbullet ~for a higher contrast limit, a first step to generate a RAP at an intermediate contrast but at a high throughput and/or a small IWA, then a second step to update the apodization on the overall pupil to increase the contrast performance: it would translate into a lower loss in throughput than with the fully redundant apodization and a lower loss in robustness than with the full aperture (non-redundant) apodization only. This aspect is similar to the apodized pupil Lyot coronagraph designs of \cite{Leboulleux2022}, which led to a higher contrast than the one accessible with the RAP only but a loss in robustness,

\quad \textbullet ~for an access to small angular separations, a first step to obtain a RAP at the target contrast but higher IWA than the target one to limit the throughput loss, then a second step to update the apodization to access a smaller IWA. Since this second step apodization is not fully redundant for low-order apodization patterns, it would also translate into a loss in robustness at small angular separations.

In addition, complex RAPs could also be considered to optimize both amplitude and phase apodization while preserving high throughputs, as already studied in \cite{Pueyo2003, Pueyo2013, Mazoyer2018, Mazoyer2018a} for two deformable mirrors or spatial light modulators, one set in a pupil plane and the second one out of pupil plane to provide both amplitude and phase apodization, combined with a classical Lyot coronagraph in \cite{Fogarty2018}. A combination of redundant apodizers with focal plane masks to obtain, for instance, apodized pupil Lyot coronagraphs could also be examined, such as for a VLT/SPHERE+-like application.

RAPs remain a robust alternative to more classical apodizers, and a passive alternative to other island effect mitigation strategies. The strategies mentioned in Section \ref{s:Low-wind effect and petaling effect} mainly rely on external components like wavefront sensors and active correction with deformable mirrors or on numerical tricks to force petal-level aberration detection. They are also limited by the capabilities of the adaptive optics correction loop (visible light sensing, loop speed...). RAPs distinguish themselves from such techniques, because they are fully passive components that can make the coronagraphic system blind to island effect aberrations: they allow us to avoid the use of a wavefront sensor and a deformable mirror dedicated to the detection and correction of the island effects.

Island effects are known as one of the main limitations affecting short angular separations, even under good observing conditions, of most of the current high-contrast imagers, and will affect upcoming giant units as well. The concept developed in this paper can be applied to the ELT as well as the TMT (see appendix \ref{s:Appendix}) and GMT architectures, or to potential upgrades of existing instruments such as SPHERE+ or SCExAO. 

In addition to these different applications, future work will address the experimental validation of the RAP concept, both in the segmentation case developed in \cite{Leboulleux2022} on the HiCAT testbed at the Space Telescope Science Institute \citep{N'Diaye2013, N'Diaye2014, N'Diaye2015b, Leboulleux2016, Soummer2018, Laginja2022} and in the island effect case developed here on the high-contrast testbed at the Institute of Planetology and Astrophysics of Grenoble. This application will preferably involve binary amplitude apodizers produced with the microdot technology \citep{Martinez2010} or with an adaptive micro-mirror array \citep[Digital Micro-mirror Device or DMD,][]{Carlotti2018a}.

\begin{acknowledgements}
This project is funded by the European Research Council (ERC) under the European Union's Horizon 2020 research and innovation programme (grant agreement n°866001 - EXACT). \\
The authors are also grateful to the referee for their pertinent and precise comments, that have helped a lot to improve the quality of this paper.
\end{acknowledgements}

\bibliographystyle{aa}
\bibliography{bib}

\begin{appendix}
\section{Application to the TMT aperture}
\label{s:Appendix}

The TMT and its 30 meter diameter aperture will have thinner spiders than the ELT \citep{Usuda2014}. However, at bad seeing conditions, it might still be subject to island effects that could, for instance, affect the high-contrast performance of the Planetary System Imager (PSI), a proposed second-generation instrument. The RAP concept could then help to mitigate such errors at large angular separations.

The aperture used in this appendix is generated by the HCIPy toolkit, which provides AO system and coronagraph simulators \citep{Por2018} and refers to the description of \cite{Jensen-Clem2021}. 

\begin{figure}
   \begin{center}
   \begin{tabular}{c}
   \includegraphics[width=7.5cm]{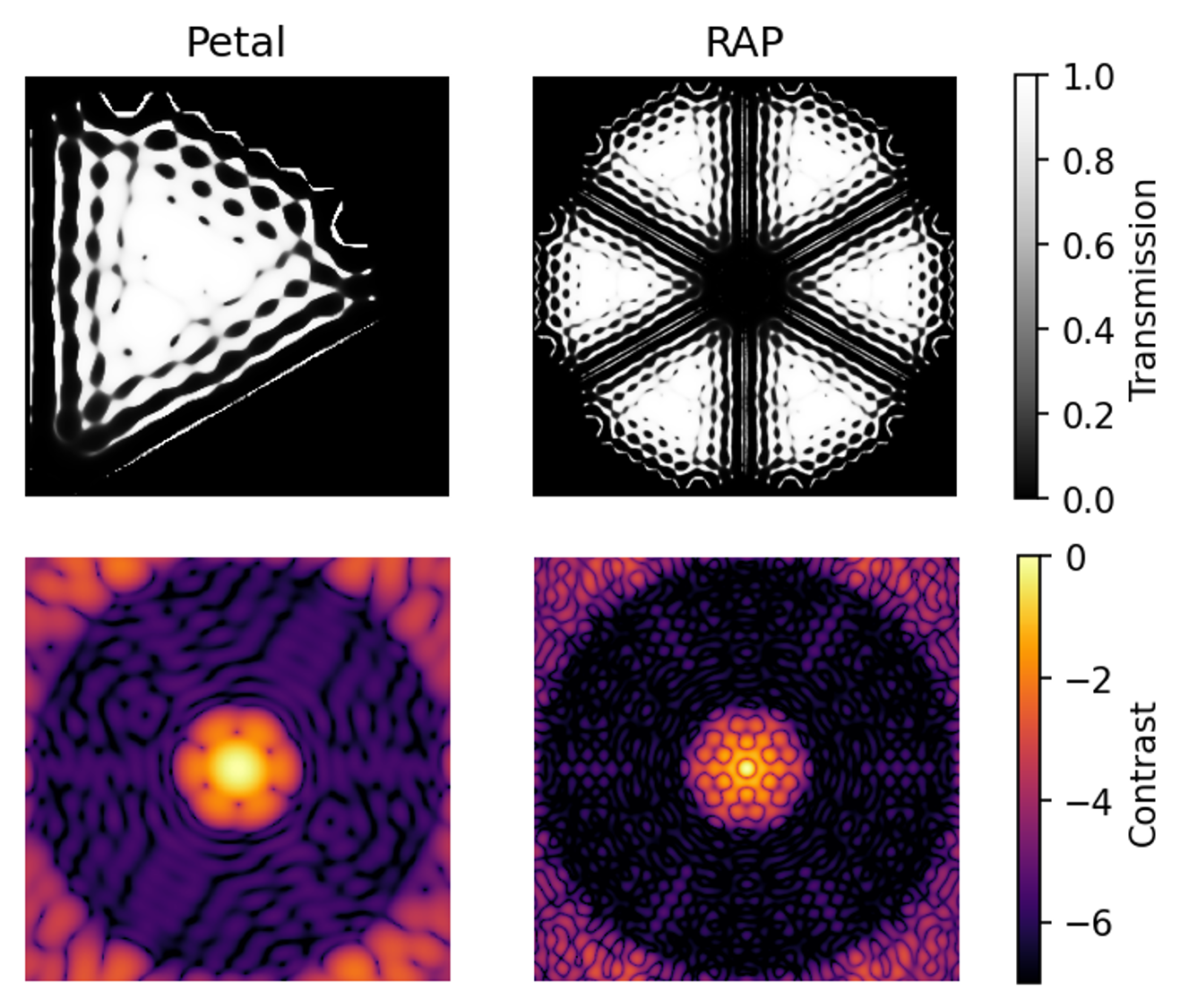}
   \end{tabular}
   \end{center}
   \caption[TMTFig1_Design] 
   { \label{fig:TMTFig1_Design} 
Amplitude RAP design for the TMT architecture: (left) for a single petal, (right) for the full RAP, with (top) the pupil transmissions from 0 (no transmission) to 1 (full transmission), aiming for a contrast of $10^{-6}$ between $8$ and $25 \lambda/D$ and (bottom) associated PSFs in logarithmic scale.} 
   \end{figure}

The SP RAP design proposed in Fig. \ref{fig:TMTFig1_Design} aims for a $10^{-6}$ contrast between $8$ and $25 \lambda/D$ and has a $22.3\%$ planetary throughput with a $47.0\%$ transmission.

\begin{figure}
   \begin{center}
   \begin{tabular}{c}
   \includegraphics[width=7.5cm]{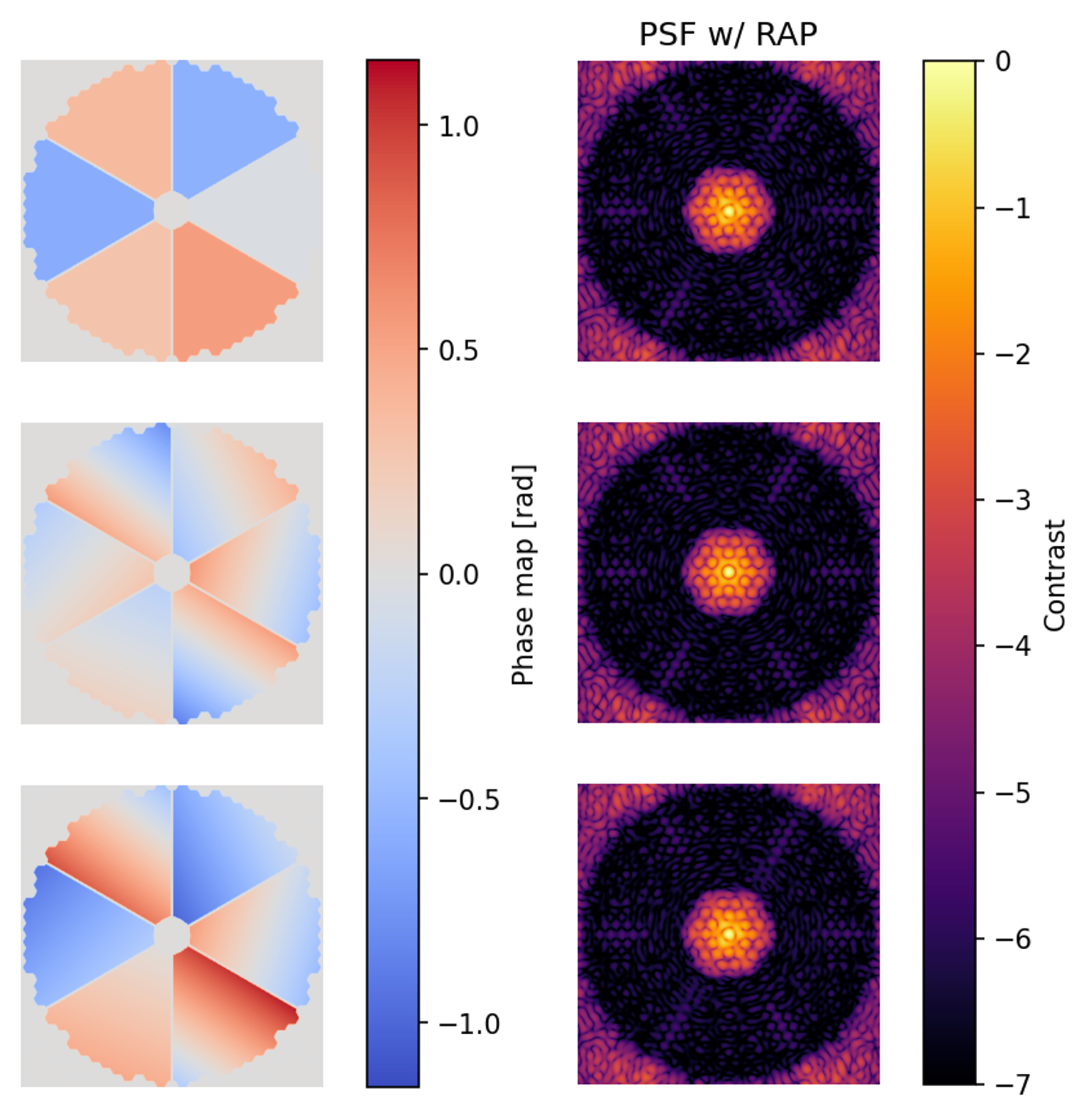}
   \end{tabular}
   \end{center}
   \caption[TMTFig2_ErrorBudget] 
   { \label{fig:TMTFig2_ErrorBudget} 
Robustness to island effect errors between petals. (left) phase maps simulated in the pupil plane in radians, (right) associated PSFs in logarithmic scale with the four designs. Three phases are considered: (top) piston only-like errors, at the level of what has been observed on SPHERE before coating, (center) tip-tilt only-like errors, also at the SPHERE level, (bottom) combined piston-tip-tilt errors from above rows.} 
   \end{figure}

Fig. \ref{fig:TMTFig2_ErrorBudget} shows the PSFs with pistons and tip-tilts between the pupil petals for RAP of Fig. \ref{fig:TMTFig1_Design}. The amplitudes of the errors have been chosen to match the values observed on the SPHERE instrument before the 2017 spider coating. The PSFs appear here robust to the considered errors.

\begin{figure}
   \begin{center}
   \begin{tabular}{c}
   \includegraphics[width=7.5cm]{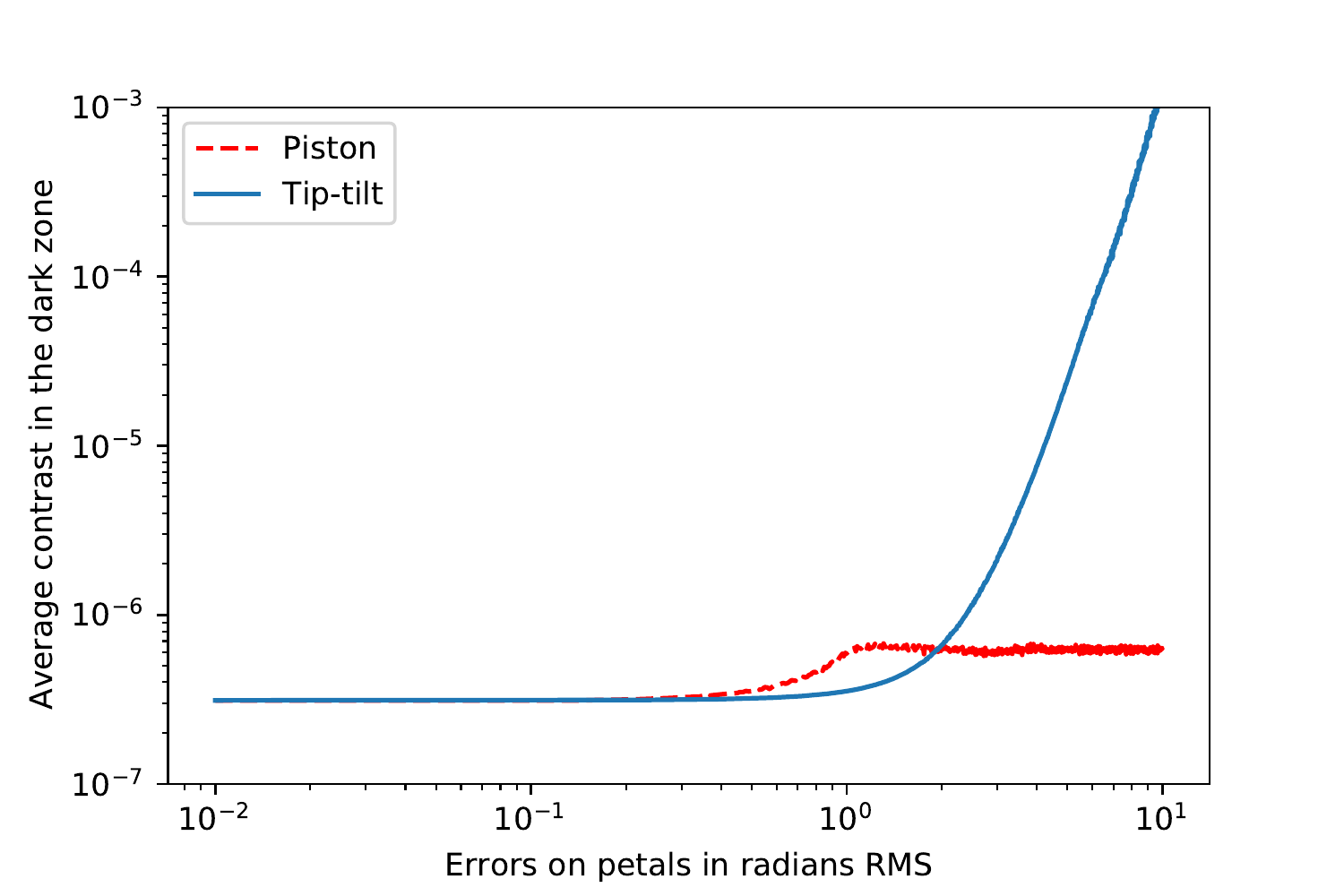}
   \end{tabular}
   \end{center}
   \caption[TMTFig3_HockeyCross] 
   { \label{fig:TMTFig3_HockeyCross} 
Evolution of the contrast with island effect errors between petals with (red) piston-like errors, (blue) tip-tilt-like errors. For each amplitude error considered here, 100 random phase maps are simulated and propagated through the coronagraph, and the average of the 100 contrasts is plotted here.} 
   \end{figure}

In Fig. \ref{fig:TMTFig3_HockeyCross}, aberrations with amplitudes between 10 mradian and 10 radians are considered, and for each amplitude 100 random phase maps are simulated and propagated through the coronagraph. For each error amplitude, the average of the 100 mean contrasts is computed and appears on this curve. It is noticeable that the RAP design shows a very stable performance for piston-like errors with a plateau at $6.2 \times 10^{-7}$ for aberration amplitudes higher than $1$ radian RMS and an average contrast in the dark region below $10^{-6}$ up to $2.4$ radians RMS of petal-level tip-tilt errors.

Since PSI will aim for higher contrasts ($\sim 10^{-8}$ at $1-2 \lambda/D$), a RAP could also be used as a first stage coronagraph and be optimized to reach higher contrasts with a slight loss of robustness at small angular separations.

\end{appendix}

\end{document}